\documentclass[a4paper,fleqn,usenatbib,twocolumn]{mnras}
\usepackage[T1]{fontenc}
\usepackage{ae,aecompl}

\usepackage{amstext}
\usepackage{url}
\usepackage{graphicx,epsfig,color,latexsym,hyperref}	
\usepackage{tabularx,natbib,bm,times}
\usepackage{subfigure}
\usepackage{amsmath}	
\usepackage{amssymb}	
\usepackage{multirow}
\usepackage{threeparttable}
\usepackage{float}
\usepackage{cases}
\usepackage{longtable}
\usepackage{color}
\usepackage{hyperref}
\usepackage{array}
\usepackage{CJK}
\usepackage{booktabs}
\usepackage{cleveref}
\usepackage{paralist}
\usepackage{ulem}
\usepackage{ulem}
\usepackage{lineno}

\graphicspath{{Figure/}}

\newcommand{\teff}{$T_{\rm eff}$}
\newcommand{\teffli}{$T_{\rm eff,Li}$}
\newcommand{\teffslam}{$T_{\rm eff,SLAM}$}
\newcommand{\teffap}{$T_{\rm eff,AP}$}
\newcommand{\teffbirky}{$T_{\rm eff,Birky}$}
\newcommand{\feh}{[Fe/H]}
\newcommand{\fehfgk}{[Fe/H]$\rm_{FGK}$}
\newcommand{\fehslam}{[Fe/H]$\rm_{SLAM}$}
\newcommand{\fehap}{[Fe/H]$\rm_{AP}$}
\newcommand{\fehbirky}{[Fe/H]$\rm_{Birky}$}

\newcommand{\loggmann}{$\mathrm {log}\,g_{Mann}$}
\newcommand{\teffmann}{$T_{\rm eff,Mann}$}
\newcommand{\logg}{$\mathrm {log}\,g$}

\newcommand{\bea}{\begin{eqnarray}}
\newcommand{\eea}{\end{eqnarray}}
\newcommand{\be}{\begin{equation}}
\newcommand{\ee}{\end{equation}}

\title[Calibration of metallicity of LAMOST M dwarf stars]{Calibration of metallicity of LAMOST M dwarf stars Using FGK+M wide binaries}
\author[D. Qiu et al.]
	{Dan Qiu$^{1,2}$, Jiadong Li$^{1,2}$,  Bo Zhang$^{1}$, Chao Liu $^{1,3,2}$\thanks{E-mail: liuchao@nao.cas.cn}, Haijun Tian$^{4,5}$, Zexi Niu$^{2}$\\
$^1$ Key Laboratory of Space Astronomy and Technology, National Astronomical Observatories, CAS, Beijing 100101, People’s Republic of China; \\
$^2$ University of Chinese Academy of Sciences, Beijing 100049, People’s Republic of China;\\
$^3$ Institute for Frontiers in Astronomy and Astrophysics, Beijing Normal University, Beijing, 102206, China \\
$^4$ School of Science, Hangzhou Dianzi University, Hangzhou, 310018, People's Republic of China.\\
$^5$Space Information Research Institute, Hangzhou Dianzi University, Hangzhou, 310018, People's Republic of China.\\
}
\date{Accepted 2023 December 18; Revised 2023 December 11; Received 2023 June 04}
\pubyear{2023}

\begin{document}

\label{firstpage}
\pagerange{\pageref{firstpage}--\pageref{lastpage}}
\maketitle
\begin{abstract}
Estimating precise metallicity of M dwarfs is a well-known difficult problem due to  their complex spectra. In this work, we empirically calibrate the metallicity using wide binaries with a F, G, or K dwarf and a M dwarf companion. With 1308 FGK+M wide binaries well observed by LAMOST, we calibrated M dwarf's [Fe/H] by using the Stellar LAbel Machine (SLAM) model, a data-driven method based on support vector regression (SVR). The [Fe/H] labels of the training data are from FGK companions in range of [-1,0.5] dex. The $T_{\rm eff}$s are selected from \citet{Li-2021}, spanning [3100,4400] K. The uncertainties in SLAM estimates of [Fe/H] and $T_{\rm eff}$ are $\sim$0.15\,dex and $\sim$40\,K, respectively, at $snri>$100, where $snri$ is the signal-to-noise ratio (SNR) at $i$-band of M dwarf spectra. We applied the trained SLAM model to determine the [Fe/H] and $T_{\rm eff}$ for $\sim$630,000 M dwarfs with low-resolution spectra in LAMOST DR9. Compared to other literature also using FGK+M wide binaries for calibration, our [Fe/H] estimates show no bias but a scatter of $\sim$ 0.14-0.18 dex. However, the [Fe/H] compared to APOGEE shows a systematic difference of $\sim$ 0.10-0.15 dex with a scatter of $\sim$ 0.15-0.20 dex. While the $T_{\rm eff}$ compared to APOGEE has a bias of 3\,K with a scatter of 62 K, it is systematically higher by 180 K compared to other calibrations based on the bolometric temperature. Finally, we calculated the $\zeta$ index for 1308 M dwarf secondaries and presents a moderate correlation between $\zeta$ and [Fe/H].
\end{abstract}


\section{Introduction}\label{sect:intro}
M dwarfs are low-mass stars located at the bottom of the main sequence in the Hertzsprung Russell diagram (HRD). They account for 70-75\% of all stars in the solar neighbourhood and dominate the nearby stellar population \citep{ Henry-2006, Bochanski-2010, Winters-2019}. The lifetime of M dwarfs in the main sequence phase exceeds the current age of the universe \citep{Baraffe-1998, Henry-2006}. Therefore, they are considered to be the excellent objects to trace the evolution history of the Galaxy \citep[e.g.,][]{West-2011, Woolf-2012, Hejazi-2015}. They also provide an important laboratory for exploring the structure and evolution of thin and thick disks in the Milky Way \citep[e.g.,][]{Ferguson-2017,Montes-2018a}. In addition, M dwarf stars have become key targets for planet hunting \citep[e.g.,][]{Mann-2011, Gillon-2017, Ribas-2018}, as their radius and mass are smaller than their solar-type counterparts. A large number of low-mass planets are expected to orbit a M-type star within its habitable zone, allowing for easier detection of low-mass exoplanets \citep[e.g.,][]{Gaidos-2007, Dressing-2013, Tuomi-2014, Kochukhov-2021}. James Webb Space Telescope (JWST) has started searching for potential life in earth-like planets orbiting M dwarfs in recent \citep[e.g.,][]{Lustig-Yaeger-2023}. Therefore, it is particularly important to obtain the accurate atmospheric parameters of M dwarfs, including effective temperature and chemical abundances.

Due to the intrinsic faintness of M dwarfs, obtaining high-resolution, high signal-to-noise ratio (SNR) spectra requires a large telescope with a long integration time. Therefore, the currently available high-resolution spectra of M dwarfs has been limited to small samples of nearby stars mainly from the Galactic disk. e.g., \citet{Rajpurohit-2014} compared 21 high-resolution spectra (R$\sim$40,000) of very low mass objects, which observed by the optical spectrometer UVES \citep{Dekker-2000} on the Very Large Telescope (VLT), with the synthetic spectra computed from the BT-Settl model \citep{Allard-2011, Allard-2012b} to determine the physical parameters of these stars. They also analyzed the molecular (TiO, VO, CaH) and atomic (Fe I, Ti I, Na I, K I) features of these high-resolution spectra of different subtypes of very low mass stars. \citet{Lindgren-2017} utilized the upgrade software package SME \footnote{http://www.stsci.edu/~valenti/sme.html} \citep{Piskunov-2017} to infer effective temperature and metallicity of 28 M dwarfs with high-resolution (R$\sim$50,000) spectra, which were obtained with the CRIRES spectrograph
at ESO-VLT \citep{Kaeufl-2004}. The precision of \teff\, and \feh\, are 100 K and 0.05 dex, respectively. \citet{Veyette-2017} determined temperature and Fe abundance of 29 M dwarfs with high-resolution spectra (R$\sim$25,000) by using the PHOENIX stellar atmosphere model \citep{Allard-2012a, Baraffe-2015, Allard-2016}. They achieved precisions of \teff\ and \feh\ with 60 K and 0.1 dex, respectively. \citet{Rajpurohit-2018} determined the stellar parameters of 292 M dwarfs by comparing the high-resolution spectra (R$\sim$90,000) observed by CARMENES \citep{Quirrenbach-2014} with the synthetic spectra from the BT-Settl atmospheric model. They found that the prominent narrow atomic lines (K I, Na I, Ca I, Ti I, Fe I, Mg I and Al I) and molecular (TiO, VO, OH, and FeH) features of the objects can be well fitted by the BT-Settl model. \citet{Woolf-2020} determined the Fe and Ti abundance of 106 M stars with spectral resolution $\sim$33,000 and $SNR>70$ by using the spectral analysis routine MOOG \citep{Sneden-1973}. \citep{Cristofari-2022} compared 12 near-IR high-resolution (R$\sim$70000) spectra of M dwarfs acquired with the SpectroPolarim{\. e}tre Infra-Rouge \citep[SPIRou;][]{Donati-2020} with two grids of synthetic spectra, PHOENIX and MARCS \citep{Gustafsson-2008} model atmospheres, respectively. The \teff\, $logg$, and $[M/H]$ of their work with internal errors  of about 30 K, 0.05 dex, and 0.1 dex, respectively. 

On the other hand, low-resolution spectra of M dwarfs can be largely collected with much more efficient observations. e.g., the Baryon Oscillation Spectroscopic Survey of the Sloan Digital Sky Survey \citep[SDSS/BOSS;][]{Dawson-2013} and the Large Sky Area Multi-Object Fiber Spectroscopic Telescope \citep[LAMOST;][]{Cui-2012, Deng-2012, Zhao-2012} have provided large amount of low-resolution spectra of M dwarfs. Consequently, developing an automatic technique to derive the precise atmospheric parameter of M dwarfs from low-resolution spectra is essential for Galactic studies. \citet{Galgano-2020} determined the effective
temperatures for 29,678 LAMOST M dwarfs with low-resolution (R $\sim$ 1800) spectra based on the supervised machine-learning code, $The \,Cannon$ \citep{Ness-2015}. And the training stellar labels they used were from the TESS Cool Dwarfs Catalog provided by \citet{Muirhead-2018}.  \citet{Du-2021} determined the atmospheric parameters for the M dwarfs with low-resolution spectra by fitting the BT-Settl model grids \citep{Allard-2011, Allard-2012b}. The intrinsic precision are 118 K and 0.29 dex for \teff\ and [M/H], respectively. \citet{Li-2021} used the stellar labels from APOGEE as the standard to calibrate \teff\ and [M/H] for $\sim$ 300,000 M dwarfs with a bias of 50 K and 0.12 dex compared to other literature. \citep{Ding-2022} fitted the LAMOST low-resolution spectra of M-type stars with the MILES spectral library \citep{Falc-2011} to determine the \teff\, $logg$ and \feh\ by using the ULySS package \citep{Koleva-2009}. The typical precision of \teff\, $logg$ and  \feh\, are 45 K, 0.25 dex, and 0.22 dex, respectively.

In recent years, machine learning techniques can efficiently process large amounts of spectral data to derive stellar parameters. \citep[e.g.,][]{Howard-2017,Ting-2019, Antoniadis-2020}. Data-driven methods have been illustrated to be promising solutions in cool star parameterization \citep{Jofr-2019}. The known information of training data sets can be transferred to the entire data sets through data-driven methods. The high performance of data-driven method in predicting stellar labels from low-resolution spectra is desirable \citep{Ho-2017}. The Cannon \citep{Ness-2015} is a widely used data-driven approach for determining stellar labels from spectroscopic data. \citet{Huang-2020} used the Cannon to determine the stellar parameters of K and M giant stars from the low-resolution spectra. The precisions of \teff\ , \logg\ , \feh, and [$\alpha$/M] of spectra with $SNR>50$ are 70 K, 0.1 dex, 0.1 dex and 0.04 dex, respectively. The Steller LAbel Machine (SLAM) \citep{Zhang-2020}, which is a data-driven method based on the support vector regression (SVR), also shows high performance in deriving stellar parameters from low-resolution spectra. e.g., \citep{Zhang-2020} determined \teff, $log\,g$, and \feh\ for $\sim$ 1 million LAMOST DR5 K giants with low-resolution spectra using SLAM. The random uncertainties of these three parameters are 50 K, 0.09 dex and 0.07 dex, respectively. \citep{Li-2021} measured \teff\ and $[M/H]$ of LAMOST M dwarfs with low-resolution spectra with SLAM and demonstrated that the \teff\ and $[M/H]$ are in agreement compare with the APOGEE results by 50 K and 0.12 dex. \citep{Qiu-2023} trained the SLAM model with LAMOST low-resolution spectra of M giant stars and the corresponding stellar labels from the APOGEE to obtain the \teff, $log\,g$, $[M/H]$, and $[\alpha/M]$. The uncertainties of \teff, $log\,g$, $[M/H]$, and $[\alpha/M]$ are 57K, 0.25dex, 0.16dex and 0.06 dex at signal-to-noise ratio ($SNR$) $>$100, respectively. 

Compared to cool-dwarf stars, the atmospheric models of F, G or K dwarfs quite accurately comply with observations, as their metallicity can be predicted from low-resolution spectra by comparing with synthetic spectra. The two members of a wide binary system are usually formed from the same molecular cloud and the metallicity of them should be same. That is, if the M dwarf star has a F, G, or K dwarf companion with known metallicity, it can be assumed that the metallicity of M dwarf is the same as that of the hotter companion \citep[e.g.,][]{Bonfils-2005, Neves-2012, Montes-2018a}. Therefore, using the F, G, or K dwarf companions with known metallicity to calibrate the metallicity of M dwarf star is a feasible technique \citep{Rojas-Ayala-2012, Mann-2013a,Newton-2014}. e.g., \citet{Rojas-Ayala-2010} made use of 17 F, G, or K dwarf companions with known metallicity to construct an empirical metallicity indicator applicable for M dwarfs with accuracy of $\sim$0.15 dex. \citet{Mann-2014} calibrated the metallicity of mid- to late-M dwarfs by F, G, K, or early-M dwarf primaries with accuracy of 0.07 dex. \citet{Porto-2017} derived the atmospheric parameters of M dwarfs from detailed analysis of F, G, or K binary companions. The internal errors are 70 K and 0.1 dex for \teff\ and \feh\, , respectively, which are calibrated by Principal Component Analysis (PCA) method. \citet{Montes-2018a} has established a sample of 192 FGK+M physically bound systems. The atmospheric parameters of M dwarf companions were calibrated by F, G, or K-type primaries. \citet{Birky-2020} used a training sample, which includes 87 M dwarfs with \feh\ labels from F, G, or K companions, to derive the \teff\ and \feh\ for 5875 M dwarfs. The prediction accuracy reaches 77 K and 0.09 dex, respectively. All the above works indicate that using F-, G-, or K-type companions with known metallicity to calibrate the metallicity of M dwarf secondaries is a common and effective way. 


In this work, we identified 1308 FGK+M wide binaries from LAMOST to calibrate the metallicity of M dwarf stars with low resolution optical spectra. This is the largest wide binary sample size used to calibrate the metallicity of M dwarf stars so far. We trained a data-driven model SLAM based on these over 1300 FGK+M wide binaries, and applied the trained model to calibrate the atmospheric parameters of  all LAMOST DR9 M dwarf stars.

This paper is organized as follows: Section \ref{sect:Data} presents the selection of FGK+M and M+M wide binary systems. In section \ref{sect:method}, we describe the data-driven method SLAM for deriving the atmospheric parameters of stars from spectrum. In Section \ref{sect:result}, we analyze the training results and validation of the trained model. The test of the trained model on LAMOST DR9 M dwarfs is presented in Section \ref{sect:Result_ana}. A discussion is shown in Section \ref{sect:discu}. Finally, we summarize the results in Section \ref{sect:conclu}.

\section{Data}\label{sect:Data}

We started with the wide binaries with a distance less than 1 kpc selected by \citet{El-Badry-2021} from the early installment of the third Gaia data release (Gaia EDR3; \citet{Gaia-2021}). The selection method similar to \citet{El-Badry-2018} and \citet{Tian-2020} is adopted. That is, limiting the projected separation between two stars less than 1 parsec, restricting the parallaxes of two components consistent within 3 (or 6) sigma. In addition, the proper motion of stars is a significant factor to distinguish genuine binaries from much more numerous pairs of unassociated stars in the field \citep{Chanam-2004}. \citet{El-Badry-2021} limited the proper motion of two stars to be similar and consistent with a Keplerian orbit. They built an initial wide binary candidates catalogue through the constraints of the above conditions. And then by counting the number of phase-space neighbours for each source of the binary candidates catalogue, they removed the candidate pairs in which either component  had neighbours larger than 30, these objects may be clusters, background pairs, or triples. The details on selection criteria of wide binaries can be referred to Section 2 in \citet{El-Badry-2021}. Finally, there left 1,871,594 wide binary candidates, which include main sequence (MS)+MS, white dwarf (WD)+MS, and WD+WD wide binaries with a small fraction of contamination, in \citet{El-Badry-2021}. Two components of a wide binary system with the brighter and fainter Gaia $G$ magnitude defined as the primary and secondary star, respectively. 

\subsection{FGK+M and M+M wide binaries}\label{sect:binary}

We cross-matched the 1,871,594 wide binary candidates with LAMOST DR9 \footnote{http://www.lamost.org/dr9/} \citep{Yan-2022}. We acquired 2453 wide binary candidates with a FGK type main sequence and a M dwarf companion (FGK+M). It is worthy to mention that there are some chance alignments among these binary candidates. In practice, we further purified the FGK+M wide binaries and obtained reliable astrometry of both components by applying following five criteria. 
\begin{enumerate}
    \item $snrg_1>15$, where $snrg_1$ represents the signal-to-noise ratio (SNR) at $g$-band of F, G, or K dwarf spectra. 
    \item $snri_2>15$, where $snri_2$ is the SNR at $i$-band of M dwarf spectra. The above two criteria are designed to ensure that the \feh\ of primary stars and the low-resolution spectra of M dwarfs are all in reasonable quality.
    \item \texttt R\_chance\_align \textless 0.1, where R\_chance\_align \footnote{It is evaluated in a seven-dimensional space as described in \citep{El-Badry-2021}.} represents the probability that a wide binary is a chance alignment. High-probability binaries will be expected to have low R\_chance\_align values. R\_chance\_align \textless 0.1 corresponds approximately to a wide binary with \textgreater 90 \% probability of being bound.
    \item $\Delta \rm rv$ (=$\lvert$ $\rm rv_1-rv_2$ $\rvert$) $<$ 20 $\rm kms^{-1}$. $\Delta$rv is the radial velocity difference between F, G, or K primaries (rv$_1$) and M secondaries (rv$_2$), the radial velocities of all stars are from LAMOST. $\Delta$rv $\rm <20\,\, kms^{-1}$ indicates that these wide binaries with high probability of being bound. 
    \item $\rm ruwe_2<1.4$, where $\rm ruwe_2$ is the Renormalized Unit Weight Error (ruwe) of the M dwarf companions in the wide binary systems. It is a quality specified by the Gaia survey \citep{Fabricius-2021}. $ruwe_2<1.4$ indicates that the M dwarf in FGK+M binary system does not have another closer companion and has a favourable astrometric observation.
\end{enumerate}
Objects that do not meet any of the above five criteria were removed. Finally, 1308 FGK+M wide binaries are left.

We also identified wide binaries with both two M dwarf companions (M+M) from \citet{El-Badry-2021} with both components observed by LAMOST. They can be used to verify the self consistency of metallicity of M dwarfs determined from the SLAM model, as described in subsection \ref{sect:self_M_M}. The selection criteria of M+M wide binaries as follows. 
\begin{enumerate}
    \item $snri_1>15$ and $snri_2>15$, where $snri\rm_1$ and $snri_2$ are the SNR at $i$-band of the primary and secondary M dwarfs, respectively.
    \item \texttt R\_chance\_align \textless 0.1.
    \item $\Delta \rm rv$ (=$\lvert$ $\rm rv_1-rv_2$ $\rvert$) $<$ 20 $\rm kms^{-1}$.
    \item $ruwe_1<1.4$ and $ruwe_2<1.4$.
\end{enumerate}
 These criteria are same as the conditions for selecting FGK+M wide binaries to ensure that both M dwarf components have reliable spectra, and these samples are likely wide binaries. Finally, we obtained 606 reliable M+M wide binaries.

\subsection{Properties of FGK+M and M+M wide binaries}\label{sect:Proper}

\begin{figure}
\centering
\includegraphics[width=0.57\textwidth, trim=1.2cm 1.5cm 0.0cm 0.0cm, clip] {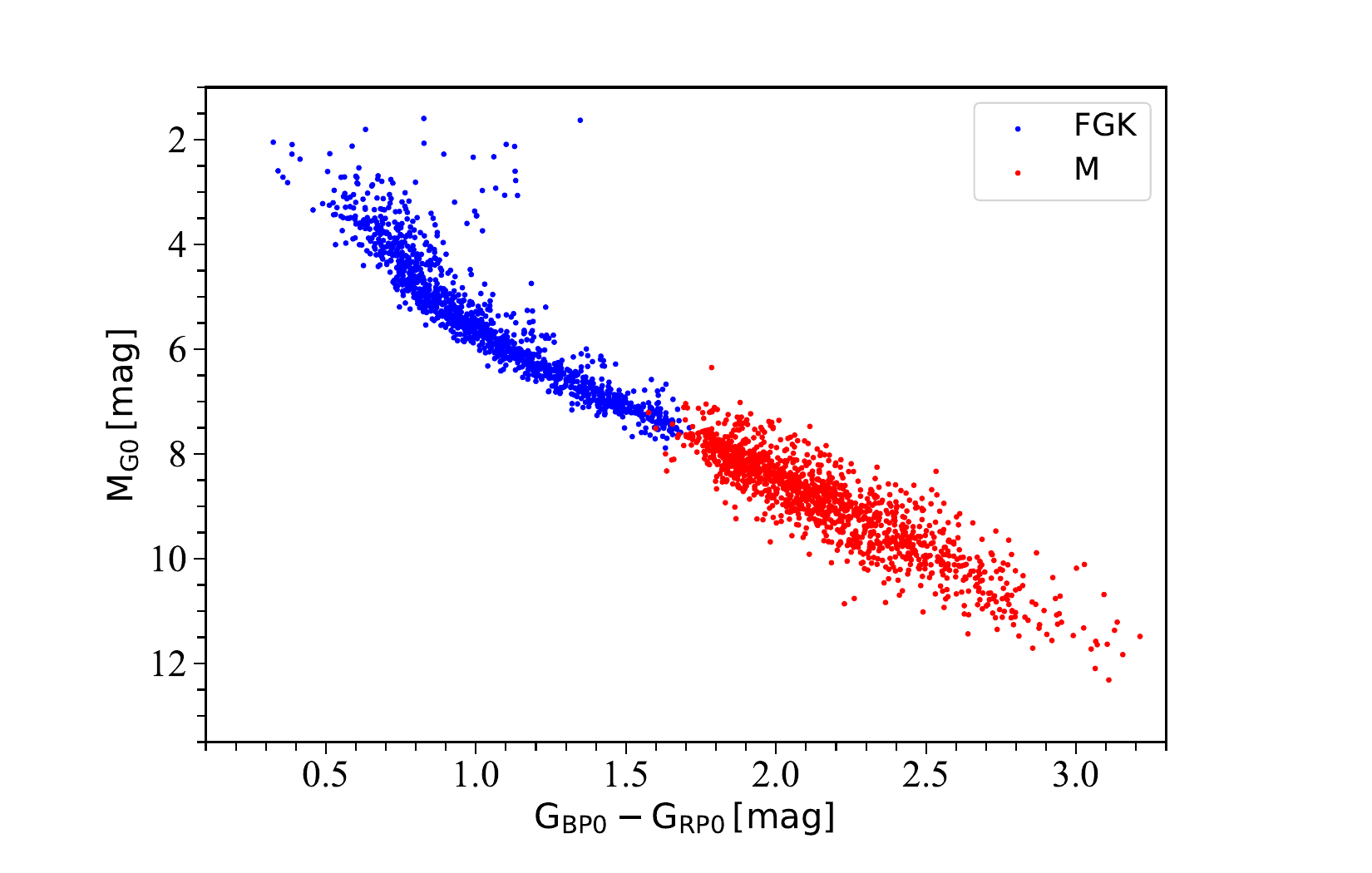}
\includegraphics[width=0.57\textwidth, trim=1.2cm 0.0cm 0.0cm 1.6cm, clip] {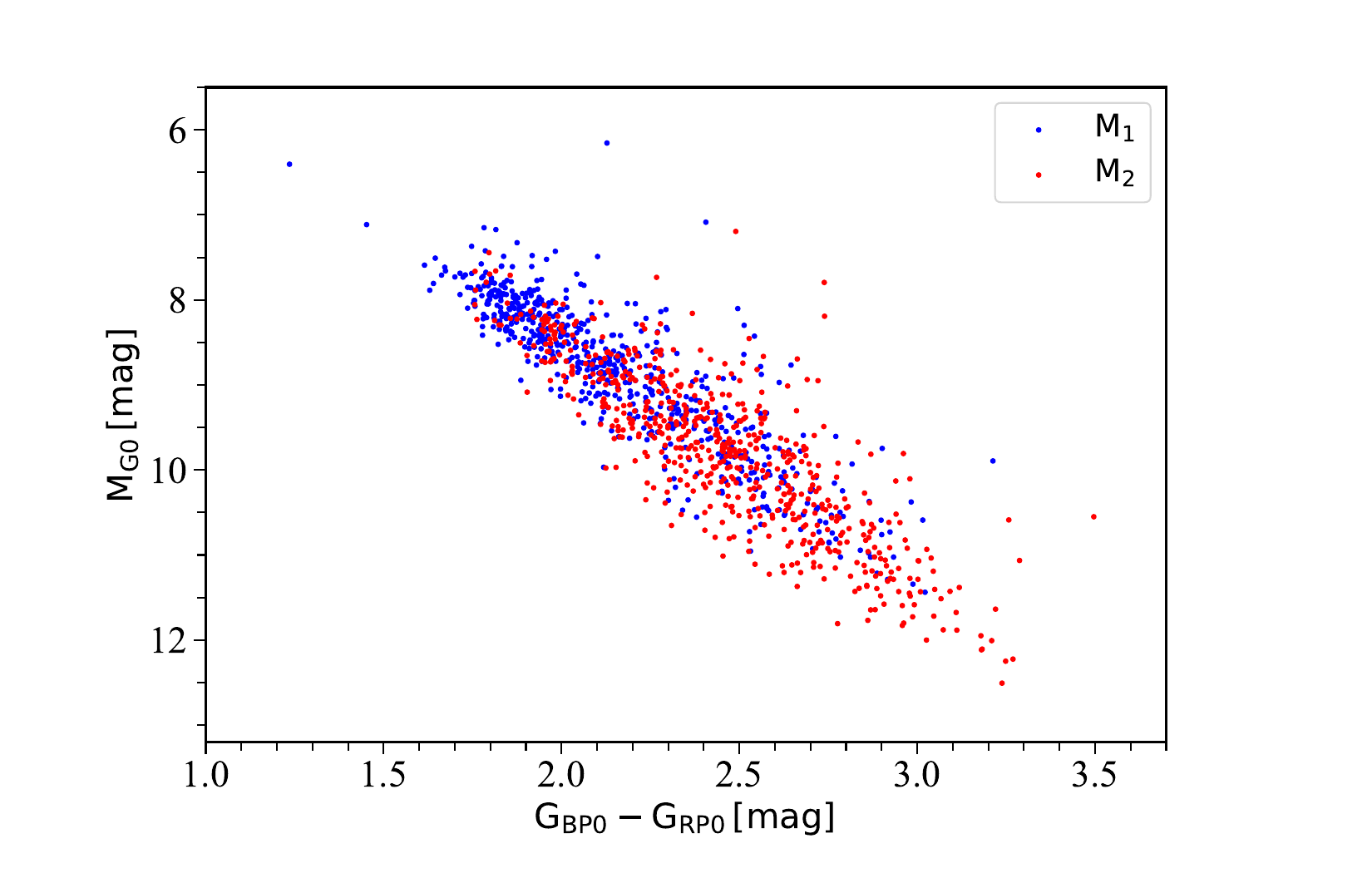}
\caption{The CMDs, i.e., \textit{Gaia} $G$-band absolute magnitude ($M_{G0}$) versus color index $G_{BP0}$-$G_{RP0}$, of the 1308 FGK+M and the 606 M+M wide binaries. Each star is corrected for the interstellar extinction. $M_{G0}$ is the extinction-corrected absolute magnitude, $G_{BP0}$ and $G_{RP0}$ are the extinction-corrected apparent magnitude. The top panel represents the CMD of the 1308 FGK+M binary systems. The blue dots are the F, G, and K dwarf primaries and the red dots indicate the M dwarf secondaries. The CMD of the 606 M+M binary systems is displayed in the bottom panel. The blue and red dots are represent the M dwarf primaries (M$_1$) and the M dwarf secondaries (M$_2$), respectively. }\label{fig:CMD_FGK_M}
\end{figure}

We obtained the reddening value $E(B-V)$ for each star from the three-dimensional dust map by \citet{Green-2019}. The extinction value can be obtained from $A_{V}=3.1*E(B-V)$. We further transferred the $A_{V}$ into the ones corresponding to the three Gaia bands, $G$, $G_{BP}$ and $G_{RP}$. The extinction coefficient of the three bands can refer to \citet{Gaia-2018a}. The color-(absolute) magnitude diagrams (CMDs) of the 1308 FGK+M and 606 M+M wide binaries are shown in Figure \ref{fig:CMD_FGK_M}. The interstellar extinction is corrected for each star. Here $M_{G0}= G+5\log(\varpi/{\rm mas})-10-A_G$, is the extinction-corrected absolute magnitude of $G$ band, where $\varpi$ is the parallax from Gaia, $A_G$ is the extinction in the $G$ band. 
The blue and red dots in the top panel represent the 1308 F, G, or K dwarf primaries and the 1308 M dwarf secondaries, respectively. Among these blue dots, a narrow branch composed of stars with $5<M_{G0}<7$ are located $\sim0.7$\,mag above the main sequence branch. These should be unresolved binary stars in triple systems, in which the F, G and K companions are binary stars. The unresolved binary \textit{primary} stars would not significantly affect the precision of the metallicity derived from spectra \citep{El-Badry-2018}. In the bottom panel of Figure \ref{fig:CMD_FGK_M}, the blue dots indicate the 606 M dwarf primaries (M$_1$) and red dots exhibit the 606 M dwarf secondaries (M$_2$).  

\begin{figure*}
\centering
\includegraphics[width=0.49\textwidth, trim=0.9cm 0.0cm 2.9cm 1.0cm, clip] {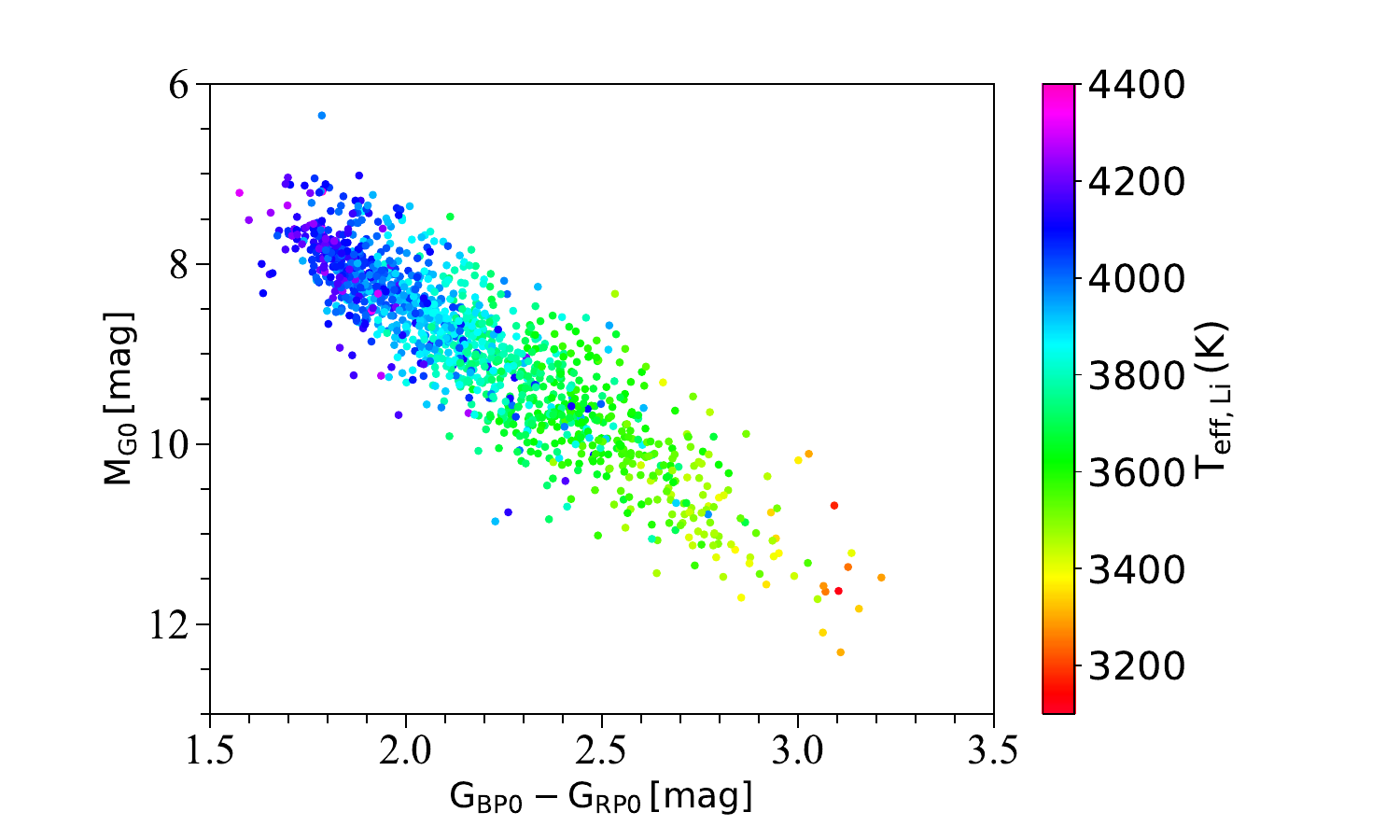}
\includegraphics[width=0.49\textwidth, trim=0.9cm 0.0cm 2.8cm 1.cm, clip] {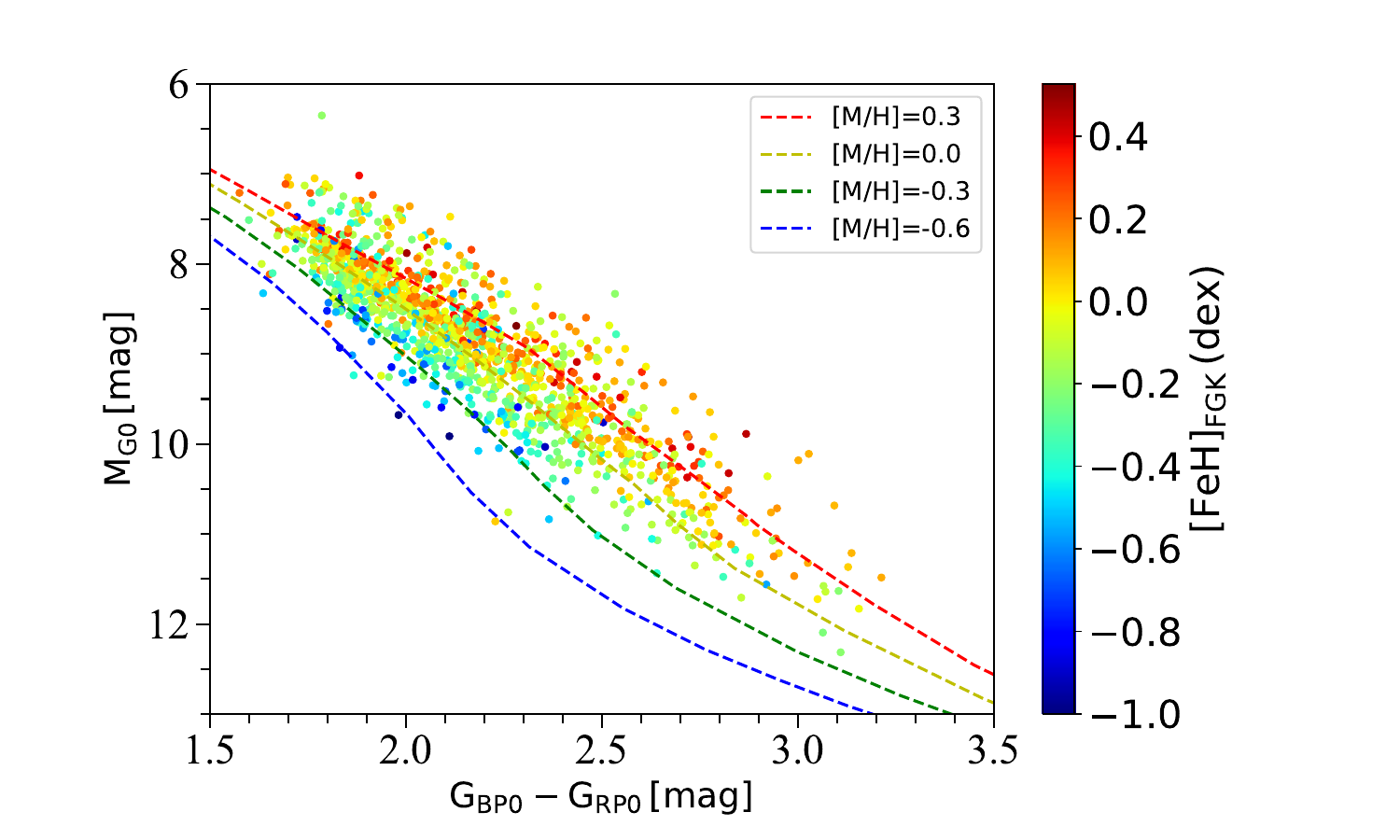}
\caption{The training sample: Two panels display the distribution of $T_{\rm eff}$ and [Fe/H] of 1308 M dwarfs in the CMD. The stellar extinction is corrected for each star. The left panel shows the CMD with color-coded effective temperature derived by the model of \citet{Li-2021}. The right panel indicates the CMD of the 1308 M dwarfs with color-coded iron abundance from F, G or K dwarf primaries. Four isochrones with different metallicity ([Fe/H]=-0.6, -0.3, 0, and 0.3 dex) at age$=10$\,Gyr from PARSEC model are shown as the blue, green, yellow and red lines, respectively.} \label{fig:BP_RP_tr}
\end{figure*}

In this work, we aim to calibrate the metallicity of M dwarfs from the F, G, or K dwarf companions. The \feh\ of the 1308 primaries (hereafter \fehfgk) are derived from the LAMOST Stellar Parameter pipeline \citep[LASP;][]{Wu-2011, Wu-2014}. Besides, we also calibrated the $T_{\rm eff}$ of the 1308 M dwarf secondaries from the trained model of \citet{Li-2021} (hereafter $T_{\rm eff,Li}$ ). The model of \citet{Li-2021} was obtained by training SLAM with LAMOST M dwarf spectra and ASPCAP stellar labels ([M/H] and $T_{\rm eff}$ ). Figure \ref{fig:BP_RP_tr} illustrates the distribution of \teff\ and \feh\ of 1308 M dwarf secondaries in the CMD. In the left panel, the colors code \teffli\ spanning in the range $3100<$\teffli$<4400$\,K. It can be seen that \teffli\ is clearly correlated with the color index. The right panel of Figure \ref{fig:BP_RP_tr} displays the CMD of the 1308 M dwarfs with color-coded \fehfgk\,ranging from -1 to 0.5 dex. The color gradient is clearly shown in the CMD: from iron-low M dwarfs (lower-left) to iron-high M dwarfs (upper-right). The four dashed lines are the theoretical isochrones from the PAdova and TRieste Stellar Evolution Code \citep[PARSEC;][]{Bressan-2012} with [Fe/H]=0.3, 0, -0.3, and -0.6\,dex at age$=10$\,Gyr, as marked by the red, yellow, green, and  blue dashed lines, respectively. It also shows that low metallicity M dwarfs are located at low-left, while high metallicity M dwarfs focus at upper-right side, which is the similar trend with the observed samples \citep{Xiong-2023, Qiu-2023}. 

\section{Method}\label{sect:method}

In this work, we used the Stellar LAbel Machine \citep[SLAM;][]{Zhang-2020}, which is a data-driven method based on support vector regression (SVR), to derive \teff\ and [Fe/H] of M dwarfs from low-resolution spectra. SVR is a robust nonlinear regression method widely used in spectral data analysis \citep{Liu-2012,Liu-2014,Lu-2015}. In previous studies, SLAM has shown good performance in determining stellar parameters from spectra \citep{Zhang-2020, Li-2021, Qiu-2023}. In next section, we described how to train a SLAM model by using 1308 FGK+M wide binaries, and then predict stellar atmospheric parameters of all LAMOST M dwarfs from the trained model. 

\subsection{Training of SLAM} \label{sect:Train_data}   
The training set includes low-resolution spectra of the low mass companion in FGK+M wide binary systems and their corresponding stellar labels (\teffli\ and \fehfgk). The \teffli\ label was derived from the trained model of \citet{Li-2021} and \fehfgk\ is from the LAMOST estimated value of the F, G, or K companions. We normalized the spectra and standardized the training set before training. We follow the preprocessing procedure of SLAM in \citet{Zhang-2020}. Firstly, we shifted the spectra back to the rest frame before training by correcting the radial velocity. And then we used a smoothing spline \citep{Boor-1978} to smooth the entire spectrum, the pixels that deviate from the smooth spectrum by a distance greater than a threshold would be excluded, e.g., 2 times the standard deviation of the residual in the wavelength bin. By smoothing the reserved pixels in the spectrum, we obtained the pseudo-continuum. The observed spectrum in the training set can be normalized by dividing its pseudo-continuum. Finally, the stellar labels and normalized spectral fluxes were rescaled, resulting in their mean and variance values are 0 and 1, respectively.

Assuming that there are $m$ spectra in the training data set and each spectrum has $n$ pixels. $F_{i,j}$ is the flux at the $j$th pixel of the $i$th normalized spectrum, 
\begin{equation}\label{eq:u}
\mu_{j}=\frac{1}{m}\sum_{i=1}^m {F_{ij}},
\end{equation}
and
\begin{equation}\label{eq:s}
s_{j}=\sqrt{ \frac{1}{m-1}\sum_{i=1}^m \big(F_{i,j} -{\mu_{j}} \big)^2\, },
\end{equation}
The normalized spectrum can be standardized via
\begin{equation}\label{eq:stan}
f_{i,j}=\frac{F_{i,j}-\mu_{j}}{s_{j}}
\end{equation}

Define $\boldsymbol{\vec{\theta}}_i$ as the stellar label vector of the $i$th star in the training set. $f_j$($\boldsymbol{\vec{\theta}}_i$)  is defined as the $j$th pixel of SLAM model output spectrum of the input stellar label vector $\boldsymbol{\vec{\theta}}_i$. The mean squared error (MSE) and median deviation (MD) can be measured by training SLAM model with a specific set of hyperparameters. Three hyperparameters, $C$, $\epsilon\,$, and $\gamma$ in SLAM need to be determined. They present the penalty level, the tube radius and the width of the radial basis function (RBF) kernel, respectively. The RBF is adopted by SLAM as the kernel. MSE and MD for $j$th pixel are defined as

\begin{equation}\label{eq:MSE}
MSE_{j}=\frac{1}{m}\sum_{i=1}^m \big[{f_j} \big(\boldsymbol{\vec{\theta}}_i\big)-{f_{i,j}}\big]^2\, ,
\end{equation}

\begin{equation}\label{eq:MD}
MD_{j}=\frac{1}{m}\sum_{i=1}^m \big[{f_j} \big(\boldsymbol{\vec{\theta}}_i\big)-{f_{i,j}}\big]\,,
\end{equation}
In principle, the smaller MSE is, the better fitting is. To avoid getting an overfitted model \footnote{Overfitting occurs when the statistical model perfectly matches its training data, in which case the algorithm cannot generalize the model to new data for prediction and data classification.} by training it on the entire training data set to seek the minimum MSE. \citet{Zhang-2020} used the k-fold cross-validated MSE (CV-MSE) and k-fold cross-validated MD (CV-MD) to evaluate $MSE_{j}$ and $MD_{j}$, i.e., the training samples are randomly divided into k subsets, and $f_{j}(\boldsymbol{\vec{\theta}}_i)$ of one subset is predicted by the trained model that depends on the other k-1 subsets of training sets. We evaluated $MSE_{j}$ and $MD_{j}$ via 5-fold cross-validation in this work. After looping over all subsets, $MSE_{j}$ and $MD_{j}$ can be calculated based on the predicted fluxes in cross-validation and the corresponding to the standardized spectral fluxes in the training set. In the case, the best hyperparameters for the $j$th pixel can be determined by searching for the lowest $MSE_{j}$ among all specific sets of hyperparameters. By doing so pixel-by-pixel, we obtained the best trained SLAM model. It is worth mentioning that the training sample includes late-type K dwarfs and M dwarfs. Consequently, the trained SLAM model can be applied to both late-type K dwarfs and M dwarfs.

\subsection{Prediction} \label{sect:Pre_data} 
According to the Bayesian theorem, given an observed spectrum, the posterior probability density function of its stellar labels is written as
\begin{equation}\label{eq:PDF}
p\big(\boldsymbol{\vec{\theta}}\mid\boldsymbol{\vec{f}}_{obs}\big)\propto p\big(\boldsymbol{\vec{\theta}}\big)\prod_{j=1}^n p\big(f_{j,obs}\mid\boldsymbol{\vec{\theta}}\big)\, ,
\end{equation}
where $\boldsymbol{\vec{\theta}}$ is the stellar label vector. $\boldsymbol{\vec{f}}_{obs}$ is the observed spectrum flux vector, in which $f_{j,obs}$ is the flux of the $j$th pixel. $p\big(\boldsymbol{\vec{\theta}}\big)$ is the prior of stellar label vector, and $p\big(f_{j,obs}\mid\boldsymbol{\vec{\theta}}\big)$ is the likelihood of observed spectrum flux of $j$th pixel ($f_{j,obs}$) given the stellar label vector ($\boldsymbol{\vec{\theta}}$). The stellar labels can be estimated by maximizing the posterior probability $p\big(\boldsymbol{\vec{\theta}}\mid\boldsymbol{\vec {f}}_{obs}\big)$. The logarithmic form of Eq (\ref{eq:PDF}) after takes a Gaussian likelihood becomes 

\begin{equation}\label{eq:likeli}
\begin{split}
\ln p\big(\boldsymbol{\vec{\theta}}\mid\boldsymbol{\vec{f}}_{obs}\big)=&-\frac{1}{2}\sum_{j=1}^n\times\frac{\big[f_{j,obs}-f_j\big(\boldsymbol{\vec{\theta}}\big)\big]^2}{\sigma_{j,obs}^2+\sigma_j\big(\boldsymbol{\vec{\theta}}\big)^2}\\
&-\frac{1}{2}\sum_{j=1}^n\times\ln\big[2\pi\big(\sigma_{j,obs}^2+\sigma_j\big(\boldsymbol{\vec{\theta}}\big)^2 \big)\big]\, ,
\end{split}
\end{equation}
 $f_{j,obs}$ and $\sigma_{j,obs}$ are the flux and uncertainty of $j$th pixel of observed spectrum, respectively. $f_{j}\big(\boldsymbol{\vec{\theta}}\big)$ and $\sigma_j(\boldsymbol{\vec{\theta}})$ are the model output spectral flux and uncertainty of the $j$th pixel corresponding to stellar label vector $\boldsymbol{\vec{\theta}}$. 

The cross-validate scatter (CV-scatter) and cross-validate bias (CV-bias) of stellar labels are defined by equations (\ref{eq:bias}) and (\ref{eq:scatter}), respectively. 

\begin{equation}\label{eq:bias}
CV\mbox{-}bias=\frac{1}{m}\sum_{i=1}^m\big(\boldsymbol{\vec{\theta}}_{i,SLAM}-\boldsymbol{\vec{\theta}}_{i}\big)\, ,
\end{equation}

\begin{equation}\label{eq:scatter}
CV \mbox{-} scatter=\frac{1}{m}\sqrt{\sum_{i=1}^m\big(\boldsymbol{\vec{\theta}}_{i,SLAM}-\boldsymbol{\vec{\theta}}_{i}\big)^2} \, ,
\end{equation}
where $\boldsymbol{\vec{\theta}}_{i,SLAM}$ and $\boldsymbol{\vec{\theta}_{i}}$ are the SLAM model predicted stellar label vector and the ground truth, respectively. According to these two equations,  the CV-scatter and CV-bias can be calculated only if the predicted stellar label vector is known. Therefore, CV-scatter and CV-bias can be considered as the standard deviation and mean deviation for the stellar labels between real and model output value, respectively. Theoretically, the smaller CV-bias and CV-scatter are, the better the prediction results of the model are. It should be noted that the CV-scatter and CV-bias are statistic of stellar labels, while the CV-MSE and CV-MD as described in Section \ref{sect:Train_data} are statistic of stellar spectra.

\section{Results}\label{sect:result}

Table \ref{col:symbol} displays the notations of the stellar labels involved in this work. We trained SLAM model with low-resolution optical spectra and corresponding two stellar parameters, \teffli\ and \fehfgk. The analysis of the trained SLAM model are displayed in the following subsections.

\begin{table} 
\centering
\caption{The notation of stellar labels.}  \label{col:symbol}
\begin{tabular}{l|c|l}
\hline\hline
Column & units & Description \\ \hline
{\fehfgk}               & dex     & \feh\ from F, G, or K companions\\
{\fehslam}               & dex     & \feh\ from SLAM \\
{\fehap}               & dex     & \feh\ from ASPCAP \\
{\fehbirky}               & dex     & \feh\ from \citet{Birky-2020}\\
{\teffslam}               & K     & \teff\ from SLAM \\
{\teffli}               & K     & \teff\ from \citet{Li-2021} \\
{\teffap}               & K     & \teff\ from ASPCAP \\
{\teffbirky}               & K     & \teff\ from \citet{Birky-2020}\\
\hline
\hline
\end{tabular}
\end{table}

\subsection{Training Results}\label{sect:train_result}

\begin{figure*}
\centering
\includegraphics[width=1\textwidth, trim=0.cm 2.5cm 0.0cm 0.cm, clip]{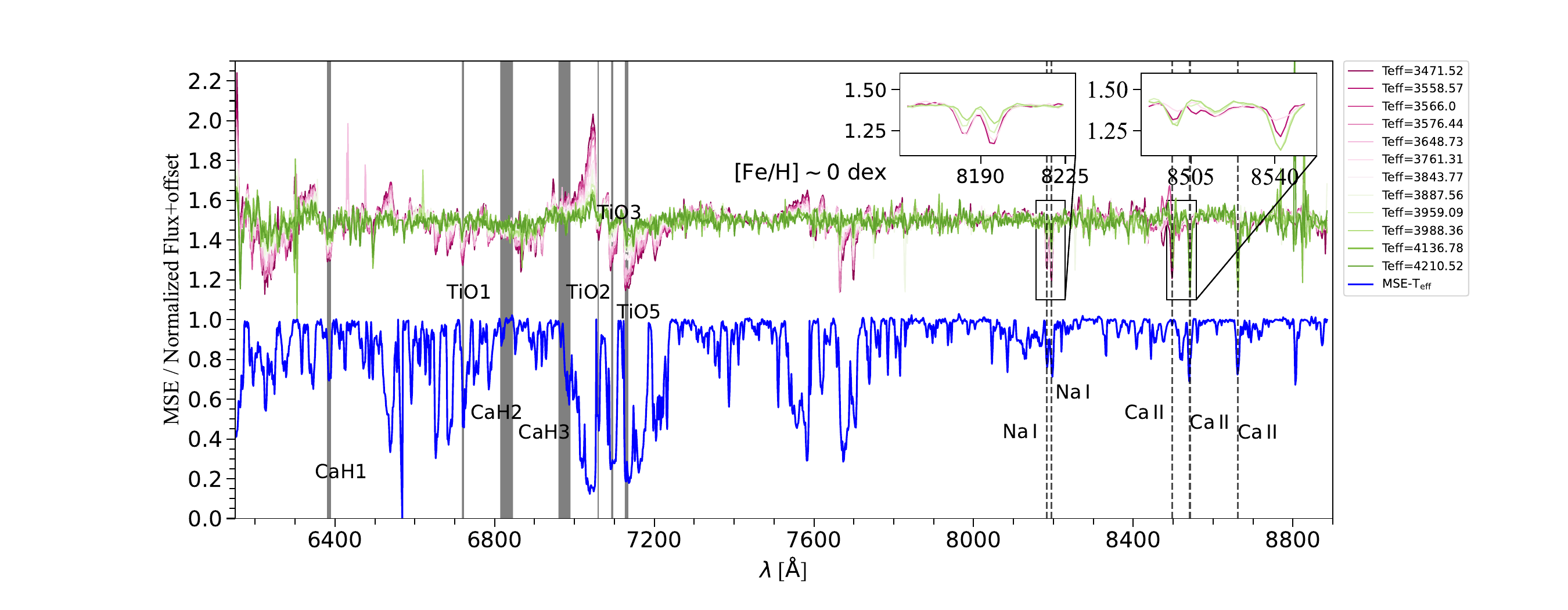}
\includegraphics[width=1\textwidth, trim=0.cm 0.0cm 0.0cm 1.cm,clip]{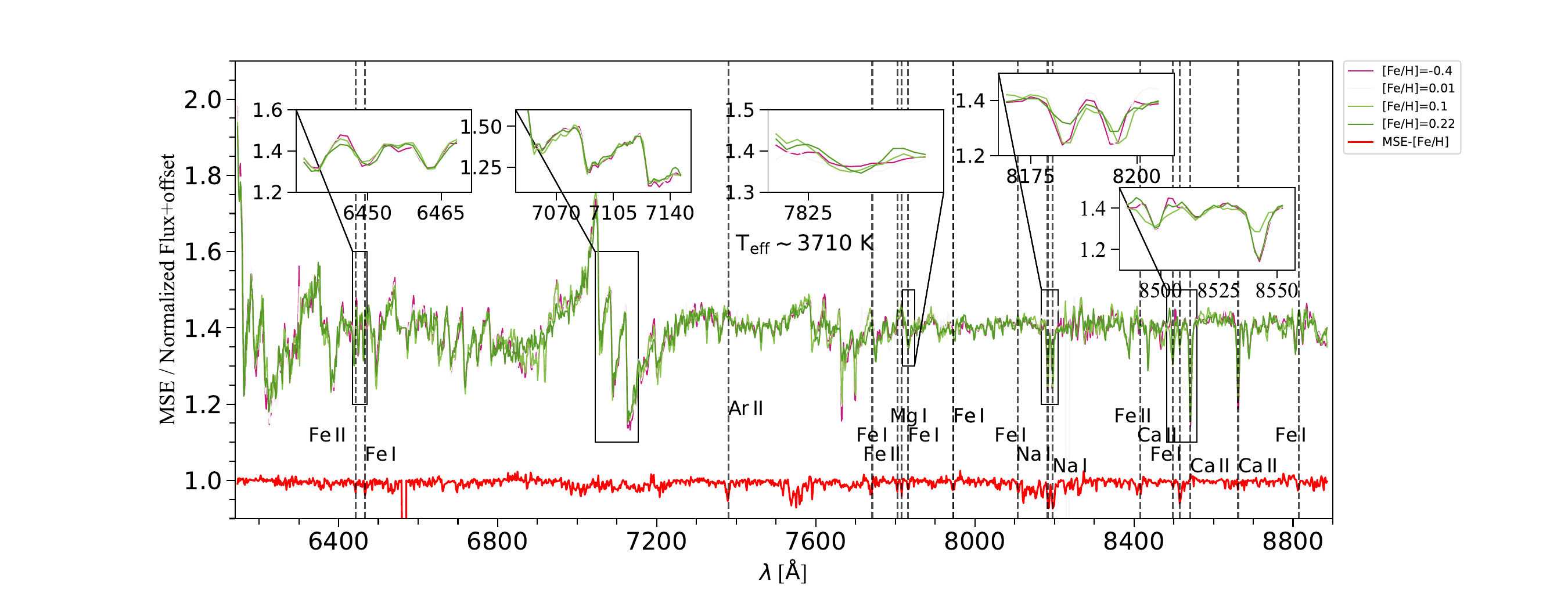}
\caption{The first panel shows the MSE-Teff (blue) and 12 spectra with the same [Fe/H]$\sim\,$0 dex but different \teff\ varying from 3471K to 4210 K (from red to green). The second panel shows the MSE-[Fe/H] (red) and 4 spectra with the same Teff $\sim$ 3710 K but different \feh\ spanning -0.4 <[Fe/H] < 0.22 dex (from red to green). The subplots in two panels are the enlarge view of flux in each spectra at the corresponding wavelengths.}\label{fig:MSE}
\end{figure*}

The percentage of variance explained (PVE) , which can be used to indicate the information content of signal in noisy data.  
\begin{equation}\label{eq:pve}
PVE_j=1-\frac{MSE_j}{{s_j}^2}
\end{equation}
where ${s_j}^2$=1 in our work. The more information is contained in the $j$th pixel about the training stellar labels, the larger value of $PVE_j$, i.e., the smaller value of $MSE_{j}$ \citep{Zhang-2020b}.
In order to derive the contribution of effective temperature and metallicity in the training model, we trained SLAM separately with \teffli\ and \fehfgk. The cross-validated MSE of \feh\ (MSE-\feh) is trained solely with \fehfgk\ and that of \teff\ (MSE-\teff) is trained separately with \teffli. As Figure \ref{fig:MSE} shows, the blue and red lines in the first and second panel represent the distribution of MSE-\teff\ and MSE-\feh\, over the wavelength range used in this work, i.e., from 6000 \AA\, to 9000 \AA , respectively. According to the value of MSE-\teff\ at each pixel,  we infer that some molecular bands and atomic lines are sensitive to effective temperature, such as the CaH and TiO bands as well as the Ca and Na lines, as emphasized by the gray bands and black dotted lines. This is consistent with some atmospheric models, e.g., the BT-Settl model  \citep{Hejazi-2020}. Besides, 12 training spectra with \feh $\sim\,$0 dex colored by \teff spanning 3471$<$\teff $<$  4210 K as displayed in the first panel.  The zoom-in subplots show the flux changes in some certain lines like Na and Ca.  The fluxes at many wavelengths with low MSE-\teff\ values varies regularly from lower to upper temperature, especially for some wavelengths in TiO band. It also demonstrates that these wavelengths are sensitive to \teff. In the bottom panel, the red line shows the distribution of MSE-\feh. It indicates that although the information about Fe abundance is weaker than that to temperature in most molecular bands and lines, some weak iron and other metal lines are sensitive to \feh. Similar to the first panel, the flux changes in the spectra with the same \teff\ $\sim$3710 K but different \feh\ ranging from -0.4 to 0.22 dex (from red to green) are displayed in the second panel. As shown in some zoom-in subplots, some certain lines like Fe I and Fe II are sensitive to \feh\ in M dwarf spectra. The values of MSE-\feh\ of these wavelengths are slight smaller than that of most other wavelengths. In the following section, we analyzed the parameter results derived from the SLAM model trained with both \teffli\ and \fehfgk simultaneously.

\subsection{Validation}\label{sect:Uncertain}

\begin{figure*}
\centering
\includegraphics[width=0.45\textwidth, trim=0.cm 0.0cm 3.0cm 1.cm, clip]{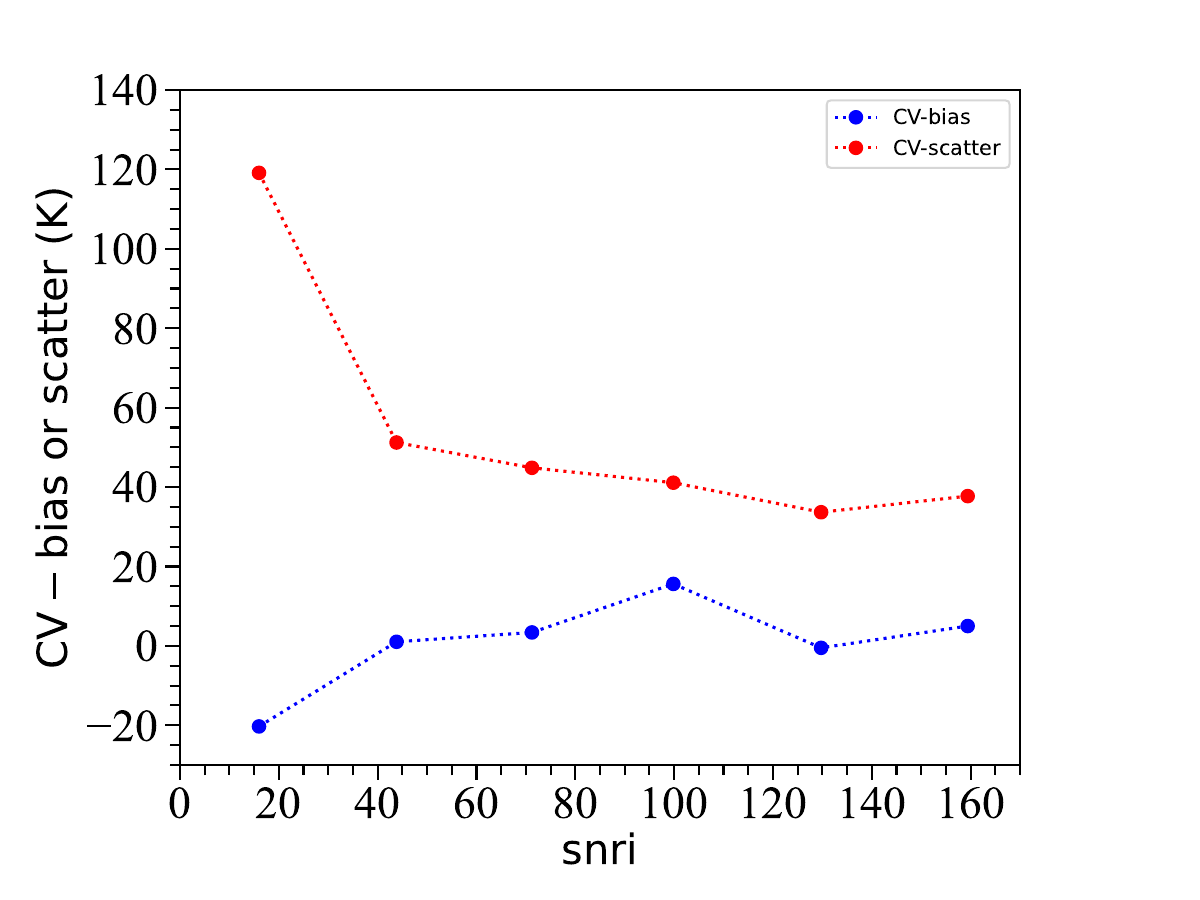}
\includegraphics[width=0.45\textwidth, trim=0.cm 0.0cm 3.0cm 1.cm,clip]{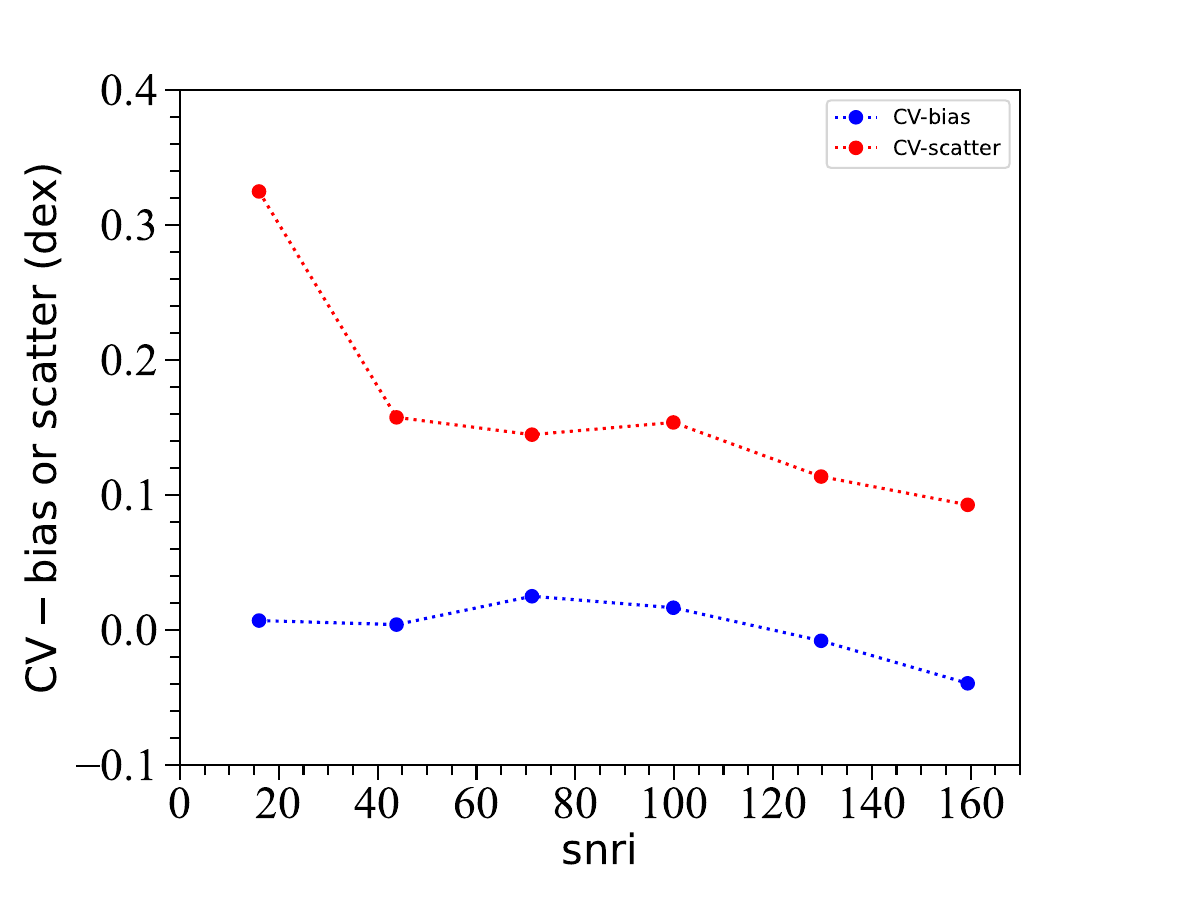}
\caption{The cross-validation (CV) bias and scatter of two stellar parameters versus the signal-to-noise at $i$ band ($snri$). In both panels, the red and blue dotted line represent the CV-scatter and CV-bias, respectively. The left panel shows the CV-bias or scatter of effective temperature changes with $snri$, the right panel represents the distribution of CV-bias or scatter of \feh\ and $snri$. The CV-scatter of temperature and \feh\ decreased with the increase of $snri$. The CV-scatter of \teff\ and \feh\ are $\sim$ 40 K and 0.15 dex at $snri\,\sim$ 100, respectively. The mean CV-bias of \teff\ and \feh\ are $\sim$5 K and $\sim$0\,dex, respectively. }\label{fig:err}
\end{figure*}

The cross-validation (CV) scatter mentioned in subsection \ref{sect:Pre_data} is usually used to quantify the precision of the predicted stellar parameters. Figure \ref{fig:err} shows how the CV-bias and CV-scatter of stellar labels change with the signal-to-noise ratio at $i$-band ($snri$). In both panels, the red and blue dotted lines represent the CV-scatter and CV-bias, respectively. The left panel of Figure \ref{fig:err} displays CV-bias and -scatter of the \teff\ versus $snri$. It demonstrates that the value of CV-scatter decreases gradually with increasing $snri$, while the CV-bias constantly fluctuates at around 5\,K. The CV-scatter of the \teff\ reach $\sim$40\,K at $snri\sim100$, and the mean CV-bias is about 5\,K. The variation of CV-bias and scatter of the \feh\ with $snri$ is drawn in the right panel and shows similar trend to the effective temperature. The CV-scatter of the \feh\ is smaller than 0.15\,dex at $snri>100$, reaching around 0.1\,dex at $snri>160$. The CV-bias of the \feh\ is almost equal to 0 at any $snri$.

\subsection{Performance}\label{sect:perform}

We adopted two methods to verify the self-consistency of the stellar parameters of M dwarfs determined by SLAM with details in subsection \ref{sect:self_cv} and \ref{sect:self_M_M}. Meanwhile, we also analyzed the stability of SLAM in predicting stellar parameters of M dwarfs, as described in subsection \ref{sect:Muli}.

\subsubsection{Self consistency in FGK+M wide binaries}\label{sect:self_cv}

\begin{figure*}
\centering
\includegraphics[width=1\textwidth, trim=0.cm 0.0cm 3.0cm 1.cm, clip]{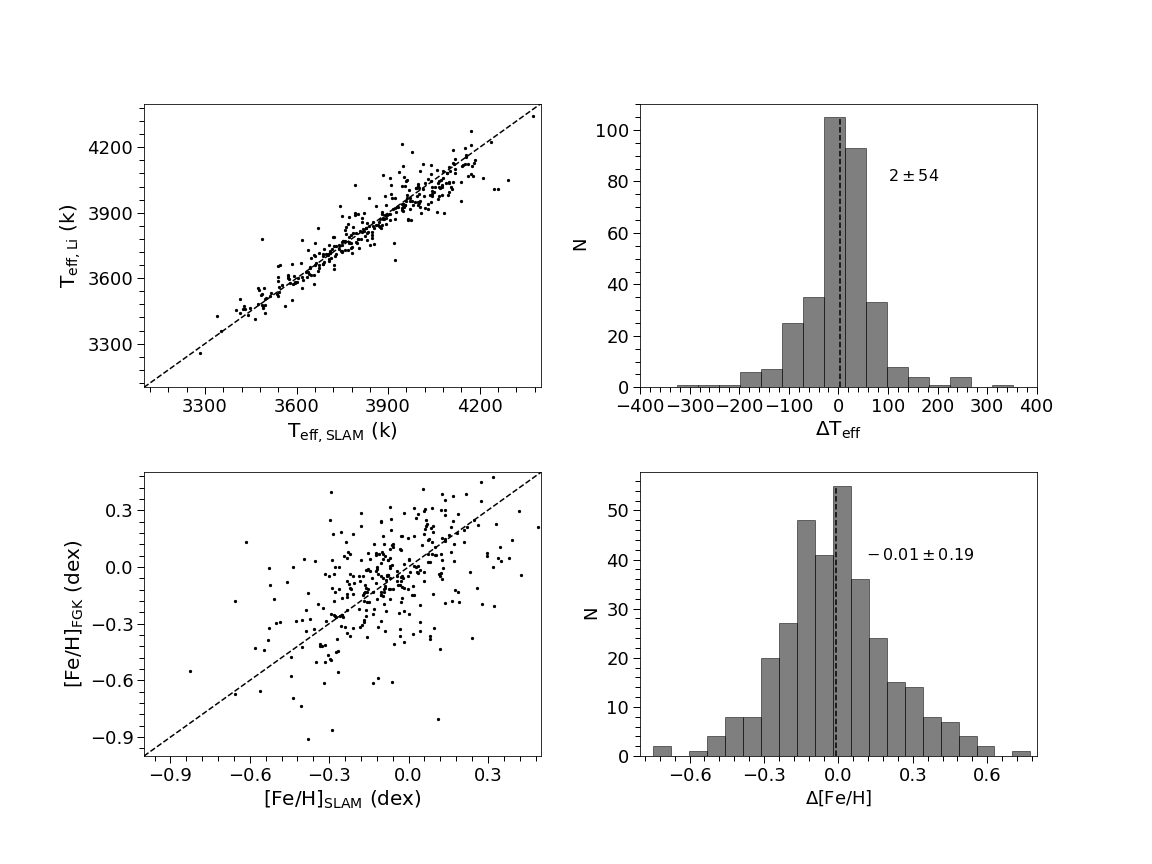}

\caption{The top-left panel is the comparison between the \teffslam\ and \teffli\ of the 308 M dwarfs. The top-right panel is the histogram of $\Delta$\teff=(\teffslam-\teffli). The comparison between the \fehfgk\ and the model predicted \fehslam\ of the 308 M dwarfs are displayed in the bottom-left panel. The bottom-right panel illustrates the histogram of $\Delta$\feh=(\fehslam-\fehfgk). The dashed black line is the one-to-one line in the top and bottom left panels. The mean value of $\Delta$\teff\ and $\Delta$\feh\ are 2 $\pm$ 54\,K and    0.01$\pm$0.19\,dex, respectively, as marked by the black dashed vertical line in the top and bottom right panels.}\label{fig:Train}
\end{figure*}

 We first checked out the self-consistency of the parameterization using the 1308 FGK+M wide binaries. First, they are randomly split out into two blocks, a training set that consists of 1000 FGK+M wide binary systems and a test set that is composed of 308 FGK+M wide binary systems. We trained the SLAM model with the 1000 M dwarf low-resolution spectra and their stellar labels (\fehfgk, \teffli). Then, we determined the metallicity (\fehslam) and effective temperature (\teffslam) for the 308 test M dwarfs with the trained model. The comparison between \teffslam\ and \teffli\ of the 308 test M dwarfs is illustrated in the top-left panel of Figure \ref{fig:Train}. The mean and scatter values of $\Delta$\teff=(\teffslam-\teffli) are 2 and 54 K, respectively, as shown in the histogram of $\Delta$\teff\ in the top-right panel. Same as the top two panels, the comparison between \fehslam\ and \fehfgk\ of the test set is displayed in the bottom two panels. The mean and scatter values of $\Delta$\feh=(\fehslam-\fehfgk) are 0.01 and  0.19 dex, respectively. It is obviously that the SLAM predicted effective temperature and metallicity are in good agreement with the ground truth, which means that the atmospheric parameters derived from the trained model are precise and reliable.

\subsubsection{Self consistency in M+M wide binaries}\label{sect:self_M_M}

\begin{figure*}
\centering
\includegraphics[width=1\textwidth, trim=0.cm 0.0cm 3.0cm 1.cm, clip]{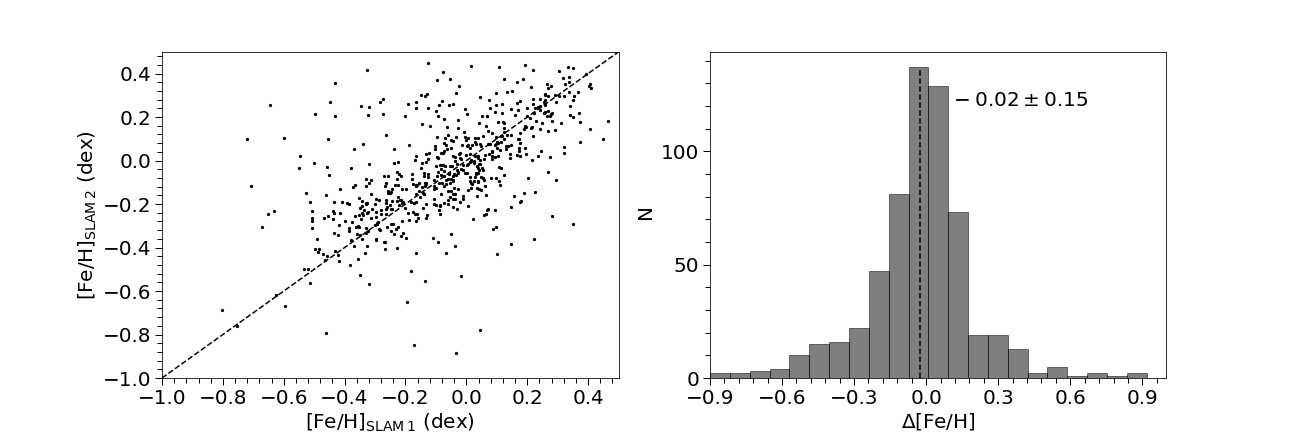}
\caption{The left panel shows the comparison of the \feh\ from the SLAM model of two components in the 606 M+M wide binaries. \fehslam$_1$ and \fehslam$_2$ represent the SLAM estimated \feh\ for the M dwarf primaries and secondaries, respectively, in the M+M wide binaries. The histogram of $\Delta$\feh (=\fehslam$_1$-\fehslam$_2$ ) of the 606 M+M binaries is presented in the right panel. The median and scatter values of $\Delta$\feh\ are 0.02 and 0.15\,dex, as marked by the black dashed line, respectively. }\label{fig:M_M}
\end{figure*}

We then selected 606 M+M wide binaries as described in Section \ref{sect:binary} to make further check on self consistency. The \feh\ of each component of the 606 M+M wide binaries are separately derived from the SLAM model. The left panel of Figure \ref{fig:M_M} displays the comparison of \fehslam$_1$ and \fehslam$_2$, where \fehslam$_1$ is the \feh\ of the M dwarf primaries and \fehslam$_2$ represents the \feh\ of the secondaries. As the distribution of $\Delta$\feh(=\fehslam$_1$-\fehslam$_2$) of the 606 M+M binaries shown in the right panel, the median value of $\Delta$\feh\ is 0.02\,dex with a scatter of 0.15\,dex. It demonstrates that \fehslam$_1$ are in good agreement with \fehslam$_2$ and indicates that the \feh\ of M dwarfs derived from the SLAM model is reasonable.

\subsubsection{Robustness}\label{sect:Muli}
\begin{figure*}
\centering
\includegraphics[width=0.498\textwidth, trim=0.cm 0.0cm 3cm 1.5cm, clip]{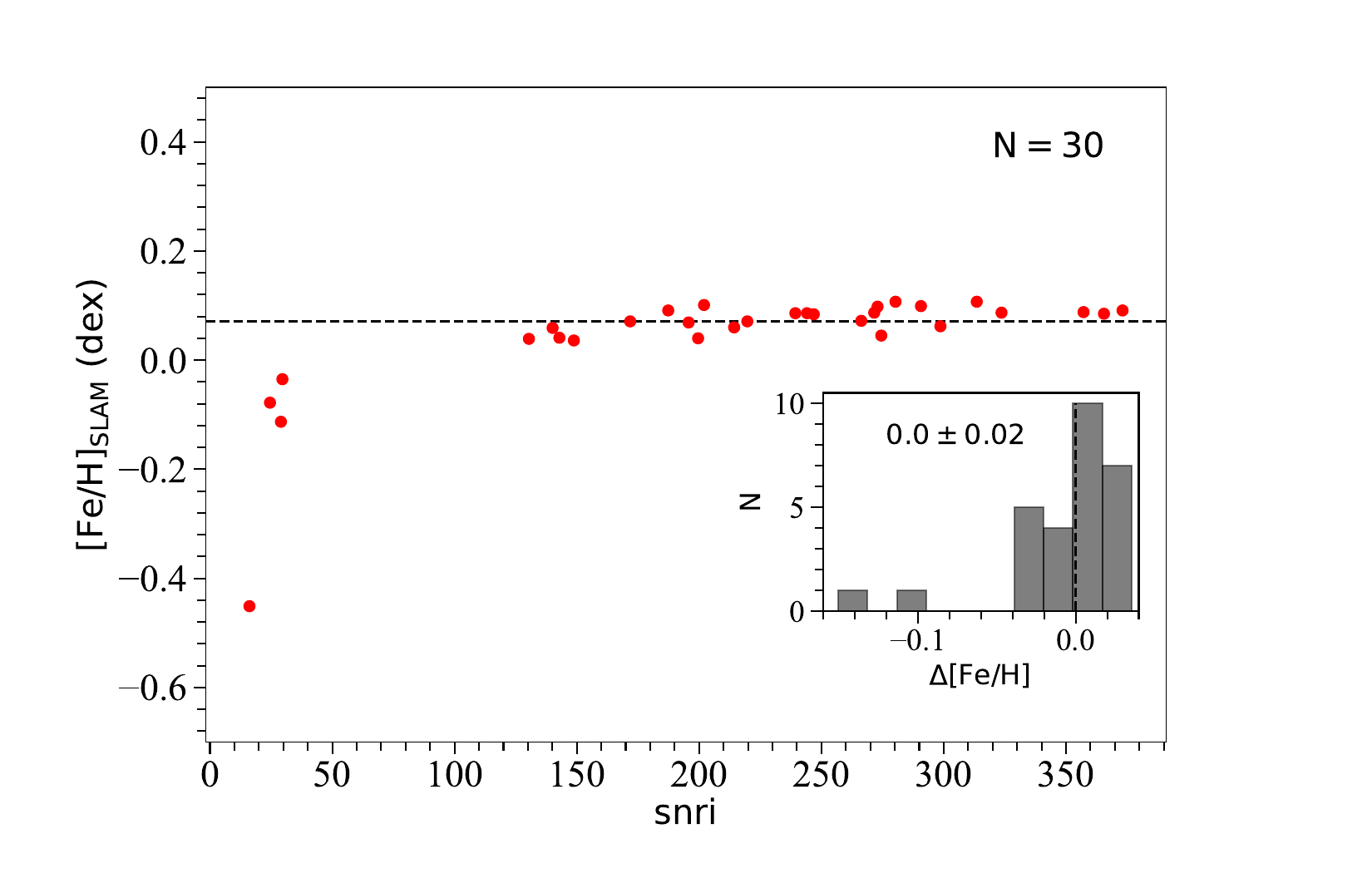}
\includegraphics[width=0.47\textwidth, trim=1.4cm 0.0cm 3.0cm 1.5cm, clip]{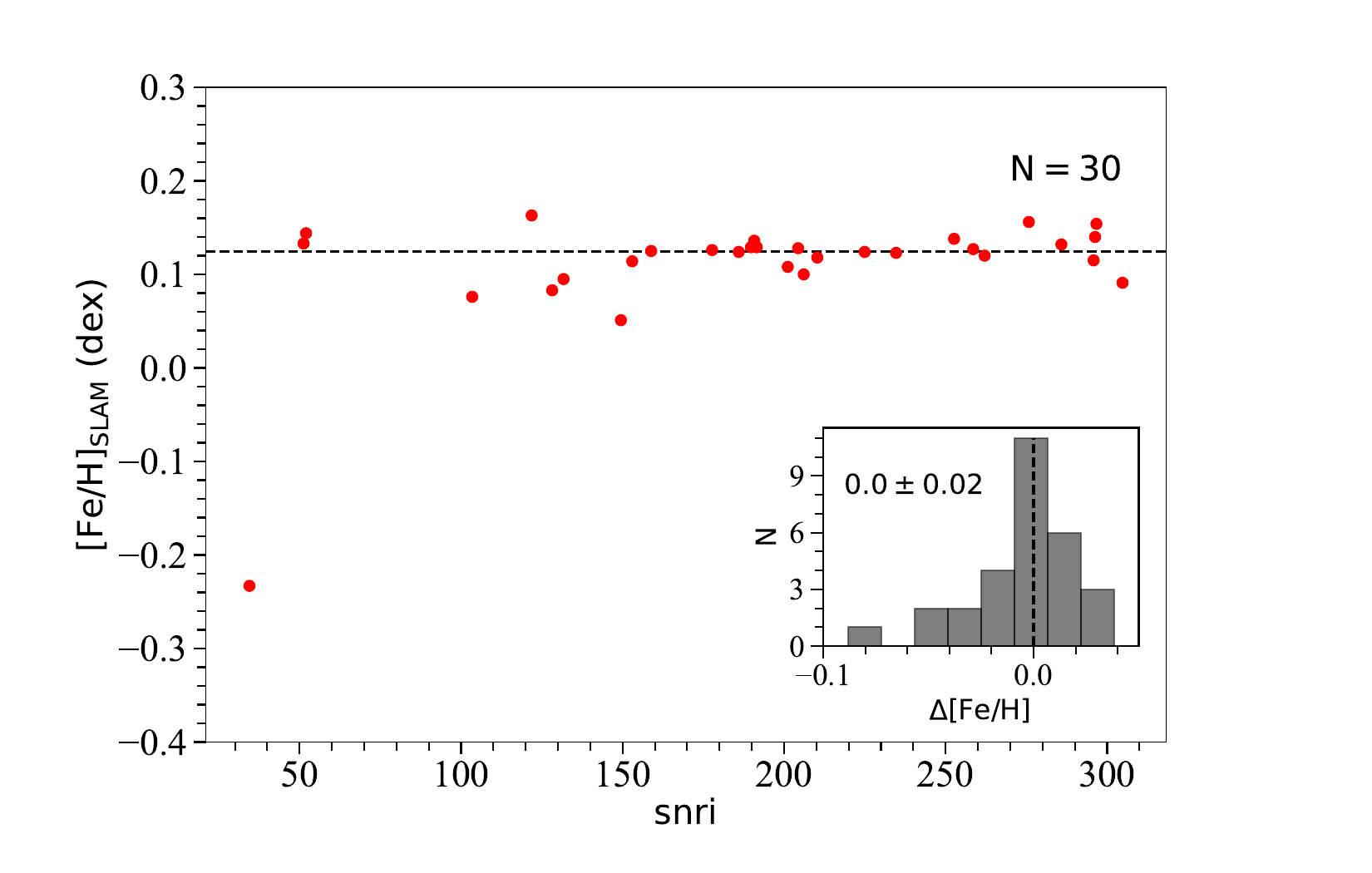}
\includegraphics[width=0.498\textwidth, trim=0.cm 0.0cm 3cm 1.5cm, clip]{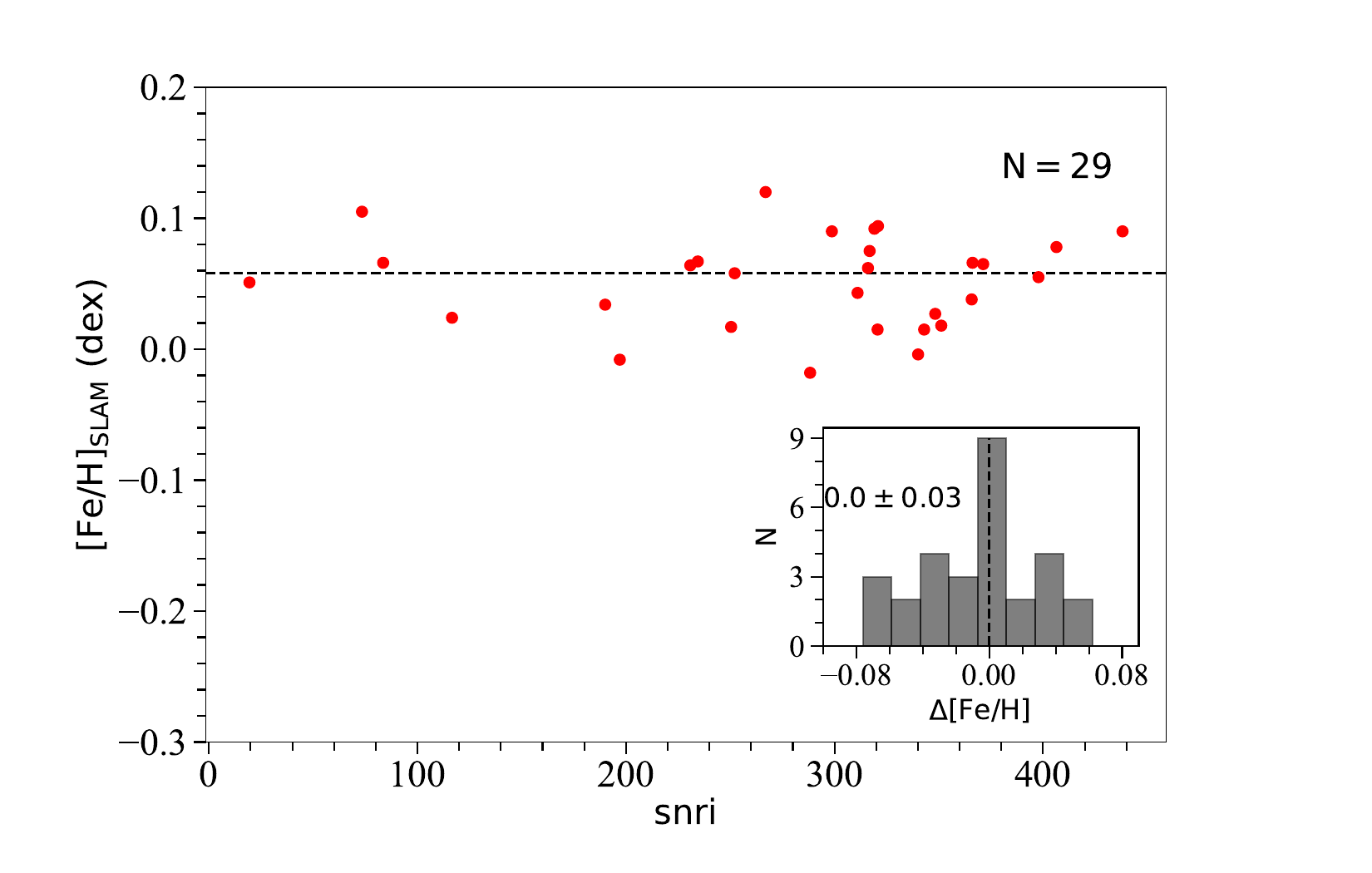}
\includegraphics[width=0.47\textwidth, trim=1.4cm 0.0cm 3cm 1.5cm, clip]{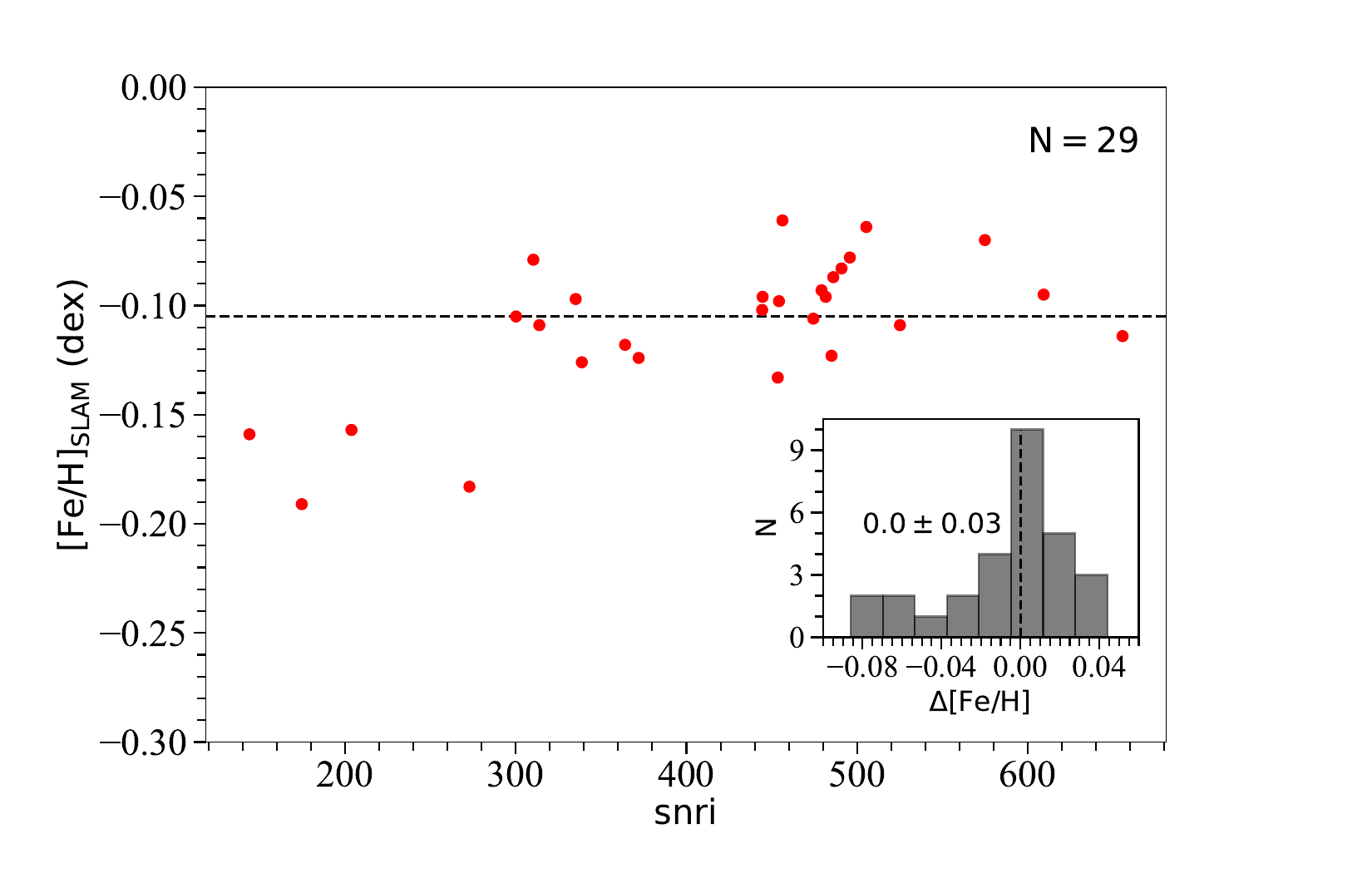}
\caption{This figure presents the \feh\ of four stars over multiple observations with different $snri$. The top two panels display the distribution of \fehslam\ and $snri$ of two stars with 30 observations, and the bottom two panels show the \fehslam\ versus $snri$ of two stars with 29 observations. The black dashed line in each panel represents the median value of metallicity (\feh$\rm_{median}$) over multiple observations. The insert sub-plot in each panel displays the histograms of $\Delta$\feh(=\fehslam$_i$-\feh$\rm_{median}$), where \fehslam$_i$ is the metallicity of the star at the $i$th observation, where $i$ is from 1 to 29 or 30. The median value of $\Delta$\feh\ is marked as the dashed line in each histogram.
}\label{fig:Mul}
\end{figure*}

We selected four stars with 29 to 30 observations, respectively, with different $snri$ in LAMOST DR9. The \feh\ of these four stars in the observations are independently measured by the SLAM model. \fehslam\ vs. $snri$ of these four stars are shown in Figure \ref{fig:Mul}. It illustrates that the \feh\ of each observation is close to the median value of \feh\ (\feh$\rm_{median}$) over multiple observations, especially for stars with high $snri$. The histogram of $\Delta $\feh(=\fehslam$_i$-\feh$\rm_{median}$) in each panel shows that the median value of $\Delta$\feh\ is around 0 with a scatter of $\sim$ 0.02-0.03 dex, where \fehslam$_i$ is the \feh\ of the star at the $i$-th observation, where $i$ is from 1 to 29 or 30. According to the metallicity results of above four stars, we find that the trained SLAM model is robust in predicting stellar parameters of M dwarf stars, especially for stars with high signal-to-noise ratios.

\section{Prediction of stellar labels for the LAMOST DR9 M dwarfs}\label{sect:Result_ana}

The large survey LAMOST has collected 11,226,252 optical spectra with low-resolution (R $\sim$ 1800) in its ninth data release. It contains more than 830,000 M-type spectra. We used the trained SLAM model to derive the atmospheric parameters from M dwarf spectra, as presented in following subsections.

\subsection{The M dwarfs}\label{sect:M_dwarfs}
\begin{figure}
\centering
\includegraphics[width=0.55\textwidth, trim=1.3cm 0.0cm 0.0cm 0.0cm, clip]{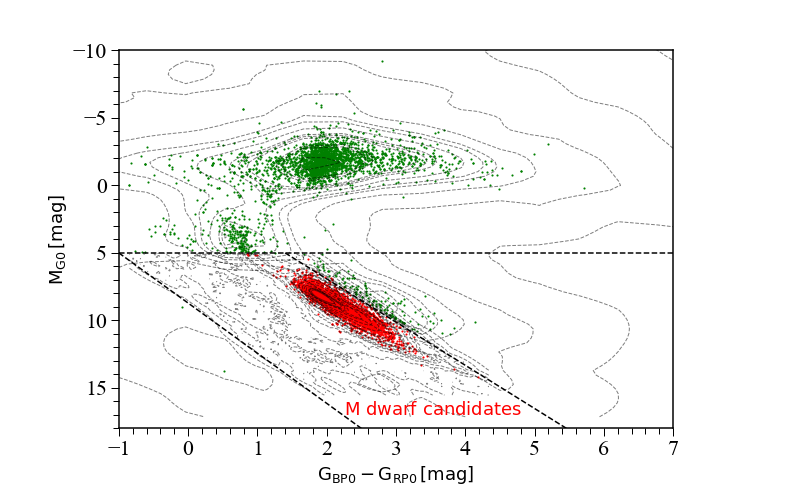}
\caption{The CMD of M-type stars from LAMOST DR9. The stellar extinction is corrected for each star. The red dots present $\sim$ 650,000 M dwarf candidates. The three black dashed lines correspond to the first three criteria in subsection \ref{sect:M_dwarfs}.}\label{fig:CMD_M}
\end{figure}

The initial catalogue of M-type stars comes from LAMOST DR9. We obtained the parallax and \textit{Gaia} three-band photometries of these stars by cross matching with \textit{Gaia} eDR3. We further purified the M dwarf samples according to their positions in the CMD. Figure \ref{fig:CMD_M} displays the CMD of $\sim$ 830,000 M-type stars. The stellar extinction correction for each star is the same as that described in subsection \ref{sect:Proper}. Obviously, there have some giants and other types of stars in the initial M-type star catalogue. We set the following four criteria to identify M dwarf candidates.
\begin{enumerate}
    \item \texttt M$\rm_{G0}$ \textgreater 5.
    \item \texttt M$\rm_{G0}$ \textless 3.71*(G$\rm_{BP0}$-G$\rm_{RP0}$)+8.71.
    \item \texttt M$\rm_{G0}$ \textgreater 3.19*(G$\rm_{BP0}$-G$\rm_{RP0}$)+0.40.
    \item \texttt ruwe \textless 1.4.
\end{enumerate}
These first three criteria are refer to \citet{Li-2021} and \citet{Birky-2020}. The criterion i) and ii) are set to remove giants and white dwarfs. Criterion iii) aims to removing the stars above the main sequence branch, which may be pre-main sequence stars, largely reddened K dwarfs with wrong extinction correction, or multiple stars. Criterion  iV) is set to remove stars with high astrometric noise or unresolved binaries. These  first three criteria are marked by three black dashed lines in Figure \ref{fig:CMD_M}. The green dots are those do not meet the above criteria. Finally, More than 650,000 M dwarf candidates left, as marked by the red dots. 

\subsection{Prediction Results}\label{sect:Pre_resu}

\begin{figure*}
\centering

\includegraphics[width=0.48\textwidth, trim=1.cm 2.cm 3.cm 1.cm, clip]{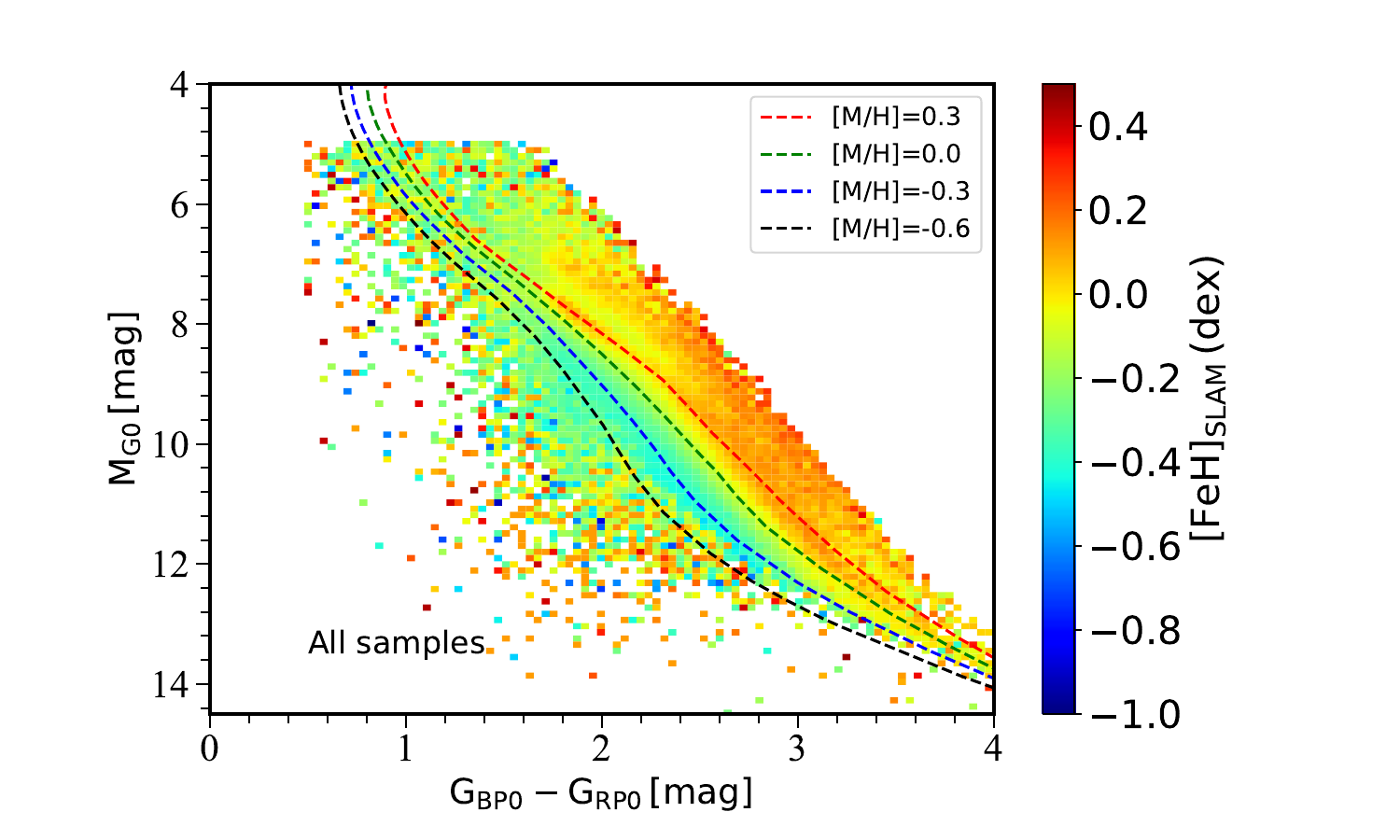}
\includegraphics[width=0.48\textwidth, trim=1.cm 2.cm 3.cm 1.cm,clip]{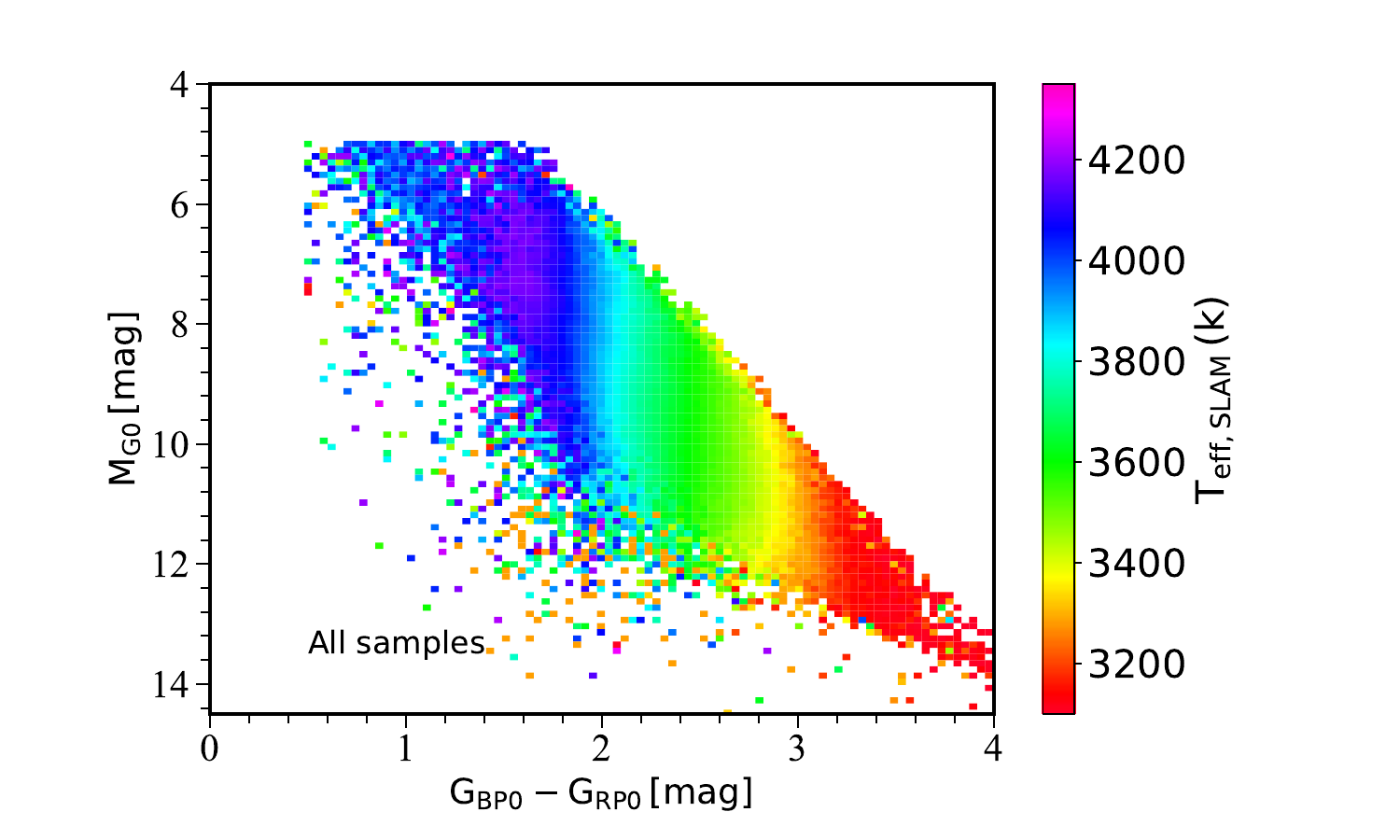}
\includegraphics[width=0.48\textwidth, trim=1.cm 0.0cm 3.cm 1.cm, clip]{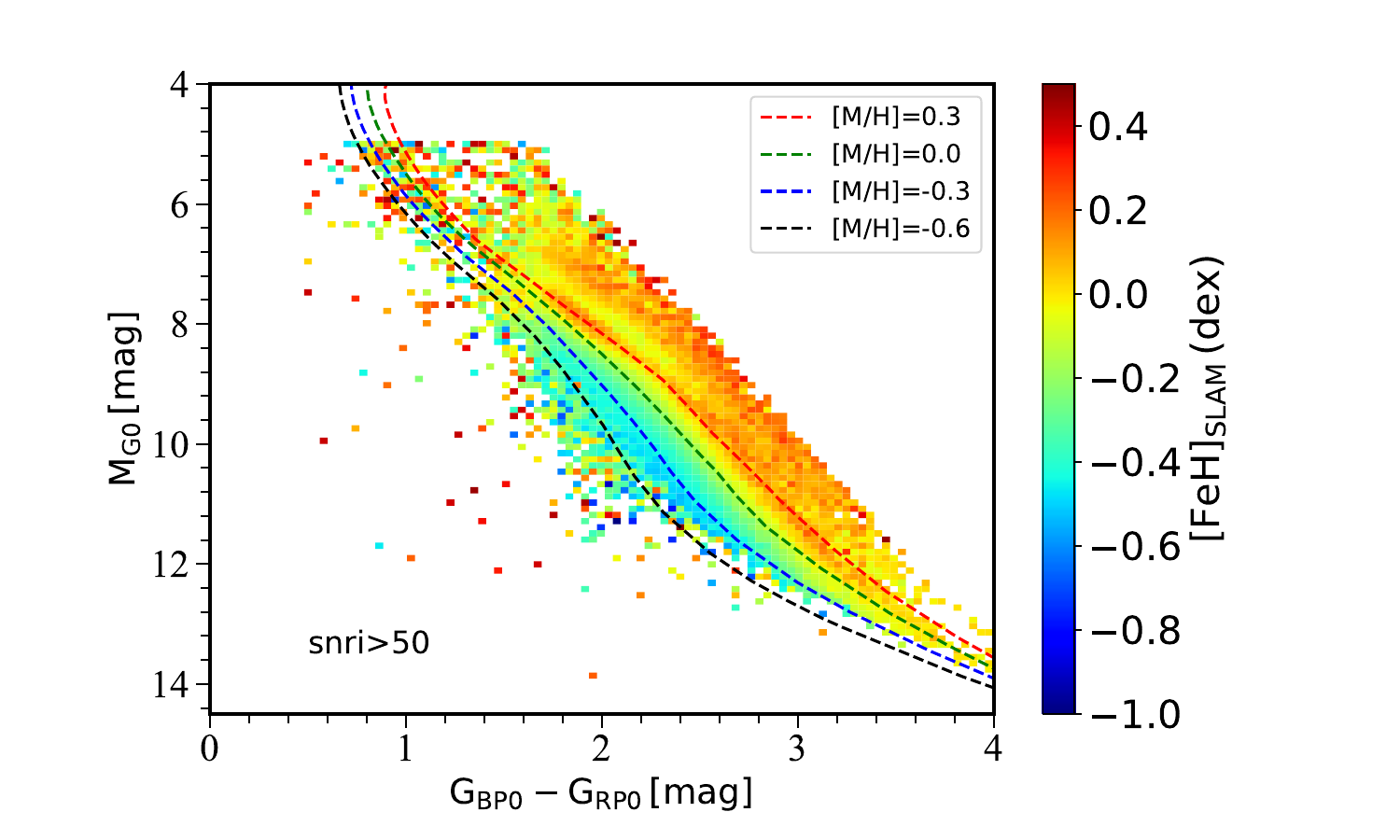}
\includegraphics[width=0.48\textwidth, trim=1.cm 0.0cm 3.cm 1.cm,clip]{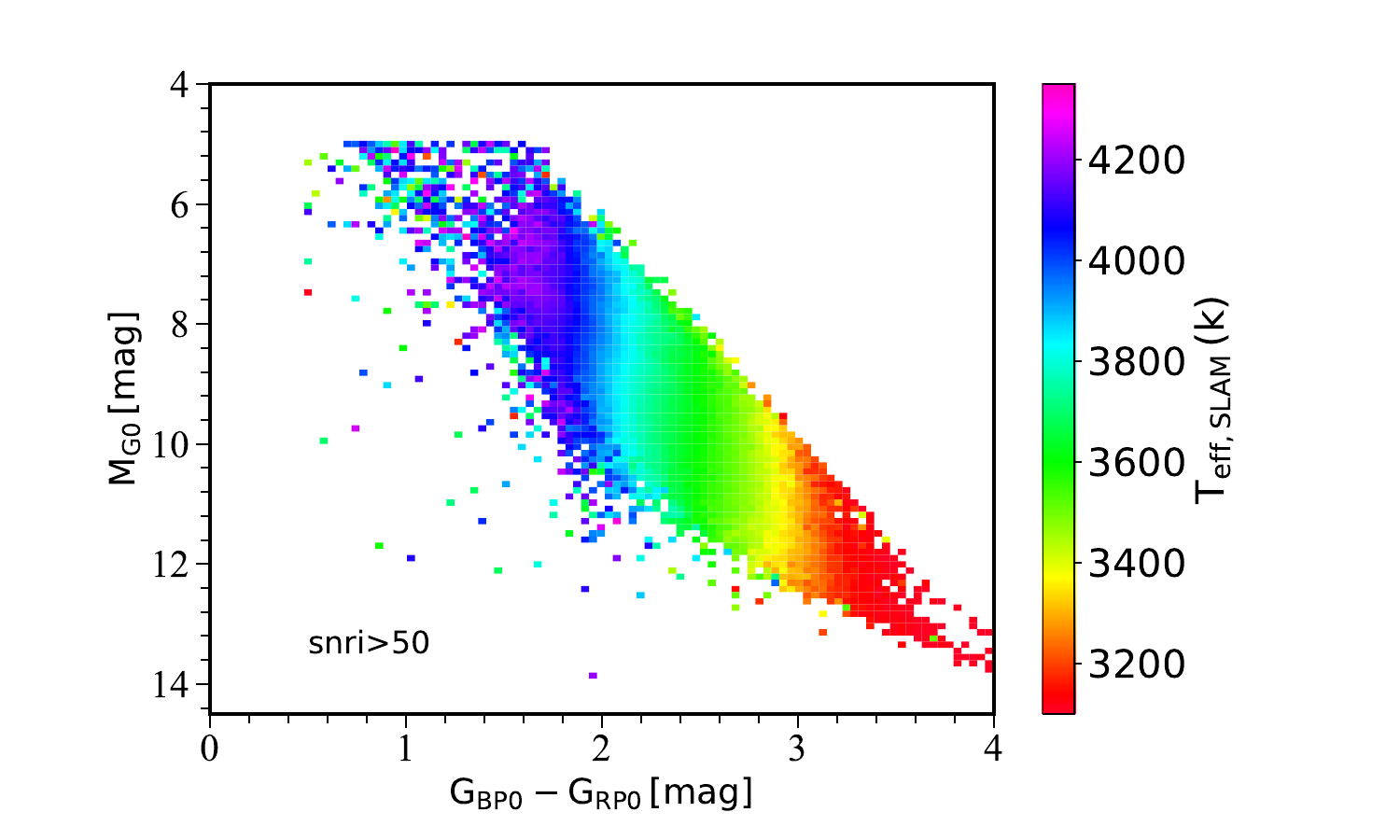}

\caption{The top two panels display the CMD of $\sim$ 630,000 M stars, and the colors code the predicted \feh\ and \teff, respectively. The top-left panel exhibits the CMD of the M dwarfs with different \fehslam\, from blue (\fehslam $<$ -0.6 dex) to red (\fehslam $>$ 0.3 dex). The four dashed lines in the top-left panel are same as those described in Figure \ref{fig:BP_RP_tr}. The CMD of $\sim$ 630,000 M dwarfs with color-coded \teffslam\ is displayed in the top-right panel. The bottom two panels are the same as the top panels but include M dwarfs with $snri\, >\, 50$. }\label{fig:params}
\end{figure*}

We derived the \feh\ and \teff\ for the $\sim$650,000 M dwarf stars from the trained SLAM model. It is noted that, due to the limitation of the SLAM model, it cannot extrapolate the stellar parameters beyond the range of the training samples. We obtained reliable stellar parameters of $\sim$ 630,000 M dwarfs after excluding stars whose predicted stellar labels that are beyond the range of the training labels. The description of the parameter catalog is shown in Table \ref{col:all}. The top two panels of Figure \ref{fig:params} show the CMDs of the $\sim$  630,000 M dwarfs. The colors code the \fehslam\ and \teffslam\ in top-left and -right panels, respectively. The top-left panel displays that the \feh\ values are located in the range of -1$<$ \fehslam $<$+0.5\,dex. 90\% of the M dwarf stars are with \fehslam $>$-0.6 dex. The four dashed lines in this panel are same as described in Figure \ref{fig:BP_RP_tr}. The colors show obvious gradient from bottom left (blue) to top right (red) in the CMD. The effective temperature of the $\sim$ 630,000 M dwarfs are in the range of 3100 $<$ \teffslam $<$ 4400\,K as displayed in the top-right panel. $\sim$ 70\% of the M dwarfs are located in between 3200 to 4000\,K. It is seen that there is a significant correlation between temperature and the color index, which is expected. The bottom two panels are the same as the top two panels but include M dwarfs with $snri\, >\, 50$. The distribution of \fehslam\ and \teffslam\ of the M dwarf stars in CMD are similar to that of the training sample, as shown in Figure \ref{fig:BP_RP_tr}. It indicates that the stellar parameters of M dwarfs derived from the SLAM model are reliable. In addition, the stellar parameters predicted by the SLAM model are consistent with the stellar evolution model, especially for stars with [Fe/H] $>$ -0.3 dex.

\subsection{Comparison of Metallicity}\label{sect:Feh_vali}
\begin{figure*}
\centering
\includegraphics[width=1\textwidth, trim=0.5cm 0.5cm 1.0cm 1.cm, clip]{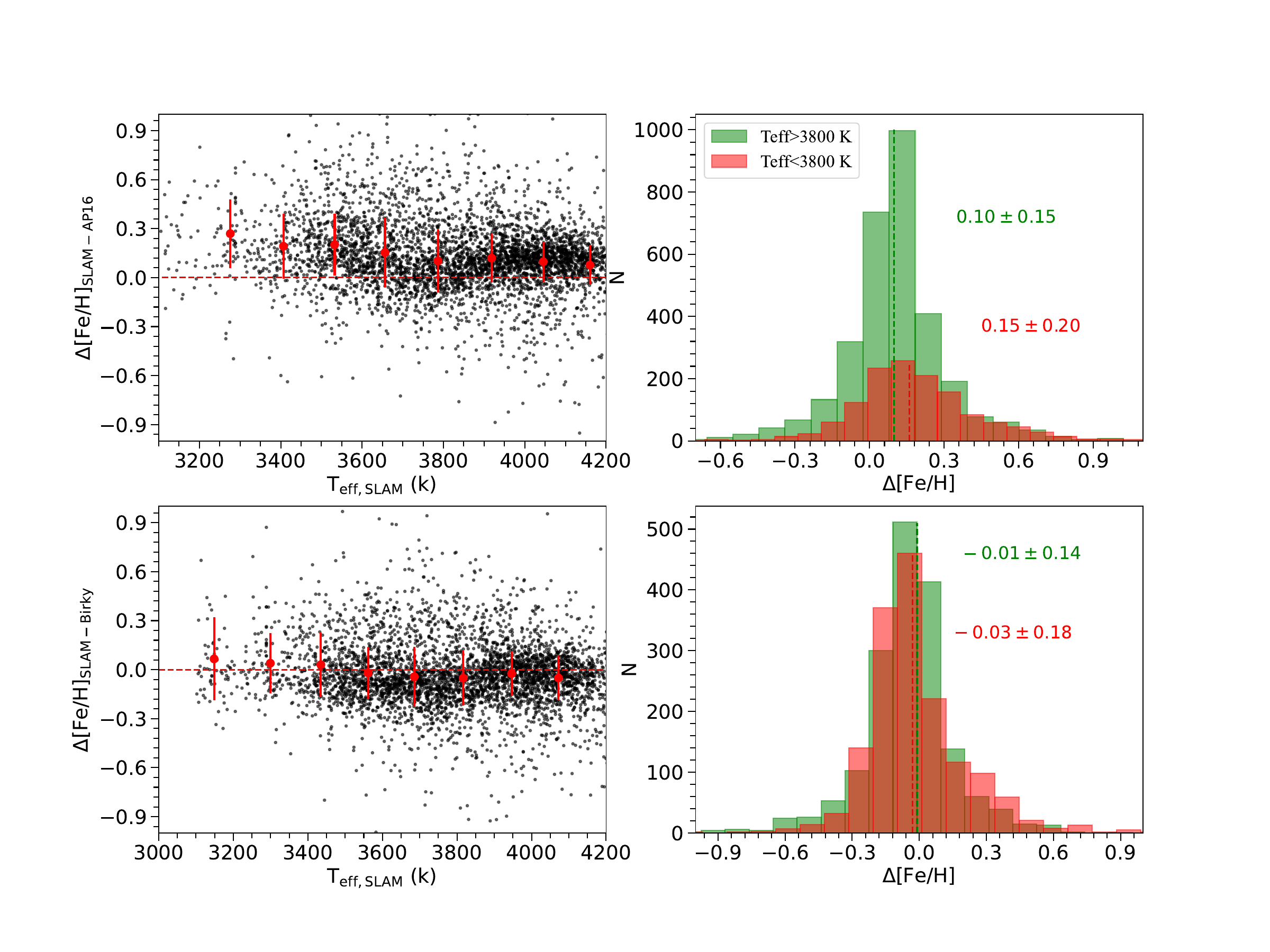}

\caption{This figure illustrates the comparison of metallicity with \fehap\ and \fehbirky. The top two panels display the $\Delta$\feh(=\fehslam-\fehap) versus \teffslam\ and the histogram of $\Delta$\feh\ of the $\sim$ 4500 common M dwarfs, respectively. The red dots and bars in the top-left panel represents the mean and dispersion values of $\Delta$\feh\ in different \teffslam\ bins. $\Delta$\feh=0 is marked as the red dotted line. The green and red histograms in the top-right panel illustrate the $\Delta$\feh\ of stars with \teffslam $>$ 3800 K and stars with \teffslam $<$ 3800 K, respectively. The mean values of $\Delta$\feh\ for the two sub-samples are exhibited by the green and red vertical dashed lines. Similar to the top two panels, the bottom two panels show the comparison of \feh\ with \fehbirky\ using the $\sim$ 3300 common stars, in which $\Delta$\feh(=\fehslam-\fehbirky). }
\label{fig:X_feh} 
\end{figure*}

As a check of the reliability of \feh\ determined by the SLAM model, we compared the predicted \feh\ in this work with two other studies. We first cross-matched the M dwarfs with APOGEE DR16 and obtained $\sim$ 4500 common stars. In Figure \ref{fig:X_feh}, the top-left panel illustrates the distribution of $\Delta$\feh(=\fehslam-\fehap) and \teffslam\,, where \fehap\ is from APOGEE. The red dots and bars indicate the mean and scatter values of $\Delta$\feh\ in different temperature bins, respectively. It is obvious that these two values are gradually increase with decreasing temperature, that is, the systematic difference between \fehslam\ and \fehap\ is larger in low temperature than that in high temperature stars. As shown in the top-right panel, the $\sim$ 4500 stars are divided into two sub-samples. The histograms of $\Delta$\feh\ of stars with \teffslam$>$3800 K and stars with \teffslam $<$3800 K are drawn in green and red, respectively. It is seen that \fehslam\ are systematically overestimated by $\sim$ 0.1-0.15 dex than \fehap, as marked by the green and red dashed lines. And the scatter of $\Delta$\feh\ for stars with \teffslam\ $>$3800 K are lower than that for stars with \teffslam\ $<$3800 K. This is likely due to the iron abundance of the cooler stars is more difficult to be determined from the complicated absorption lines. 
The systematic difference between \fehslam\ and \fehap\ may be related to many factors, such as the spectral resolution, wavelength ranges and methods used in this work are different from those used in APOGEE. In addition, the opacity and incomplete atomic or molecular lines in the atmospheric model MARCS, which is used by the APOGEE pipeline, may also contribute to the difference.
We then verified the rationality of \feh\ in this work by comparing with \citet{Birky-2020}, who trained a two-parameter model (using \textit{The Cannon}, \citet{Ness-2015,Casey-2016,Behmard-2019} ) with high-resolution $H$-band spectra of 87 M dwarfs in FGK+M wide binaries. The training labels spanning from \teff=2860 to 4130\,K calibrated with bolometric temperatures and from \feh=-0.5 to 0.5\,dex calibrated with F, G, or K-type dwarf companions. We obtained $\sim$3300 common M dwarfs both in our work and \citet{Birky-2020}. Similar to the top two panels, the bottom two panels exhibit the comparison of \feh\ with \citet{Birky-2020}. It is seen that the mean values of $\Delta$\feh\ are close to 0, especially in high temperature bins. It indicates that \fehslam\ is consistent with \fehbirky, which is expected since that \fehbirky\ is also calibrated by F, G, or K dwarf companions. The scatter of $\Delta$\feh\ also tends to increase with decreasing temperature.

\subsection{Comparison of effective temperature }\label{sect:teff_vali}
\begin{figure*}
\centering
\includegraphics[width=1\textwidth, trim=0.cm 0.0cm 2.9cm 1.0cm, clip]{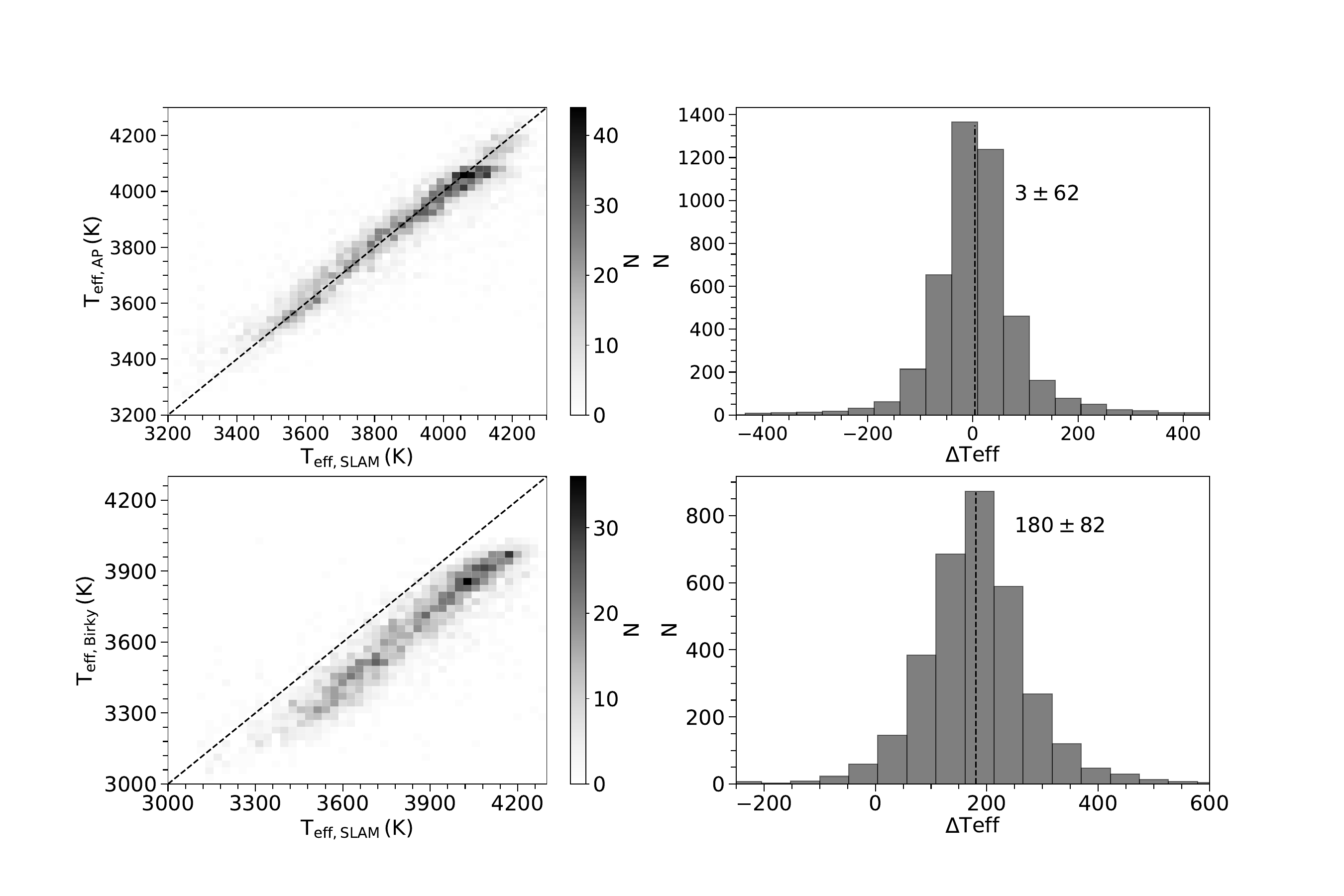}

\caption{The top-left panel shows the comparison of \teffslam\ and \teffap\ of the $\sim$4500 common M dwarfs. The black dashed line is the one-to-one line. The histogram of $\Delta$\teff=(\teffslam-\teffap) is displayed in the top-right panel. The mean value of $\Delta$\teff\ is marked as the black dashed line. The bottom two panels are similar to the top two panels. They exhibit the comparison of \teffslam\ with \teffbirky\ with the $\sim$ 3300 common M dwarfs, in which $\Delta$\teff=(\teffslam-\teffbirky).  }\label{fig:X_teff}
\end{figure*}

We take a comparison with the effective temperatures from APOGEE DR16 and with those from \citet{Birky-2020}. The \teffslam\ vs. \teffap\ of the $\sim$ 4500 common M dwarfs is displayed in the top-left panel of Figure \ref{fig:X_teff}, where \teffap\ is the effective temperature of stars from APOGEE. The histogram of $\Delta$\teff=(\teffslam-\teffap) is shown in the top-right panel. The color bar represents the number of stars in each \teffslam\ vs. \teffap\ bin. The median and dispersion values of $\Delta$\rm \teff\ are 3 and 62\,K, respectively. It demonstrates that \teffslam\ and \teffap\ are in good agreement. This is expected as the training stellar label \teffli, which is directly from \citet{Li-2021}, was estimated by using the effective temperatures from APOGEE as the training labels. The bottom-left panel of Figure \ref{fig:X_teff} displays \teffslam\ versus \teffbirky\ of the $\sim$ 3300 common stars. According to the distribution of $\Delta$\teff=(\teffslam-\teffbirky) as displayed in the bottom-right panel, we find that the temperatures of \citet{Birky-2020} are underestimated by 180$\pm$82\,K. This is consistent with the comparison between \teffbirky\ and \teffap\ in Figure 8 of \citet{Birky-2020}.  

\section{Discussion}\label{sect:discu}

\subsection{Metallicity index $\zeta_{\rm TiO/CaH}$}\label{sect:zata}

\begin{figure}
\centering
\includegraphics[width=0.55\textwidth, trim=0.cm 0.0cm 0.0cm 0.cm, clip]{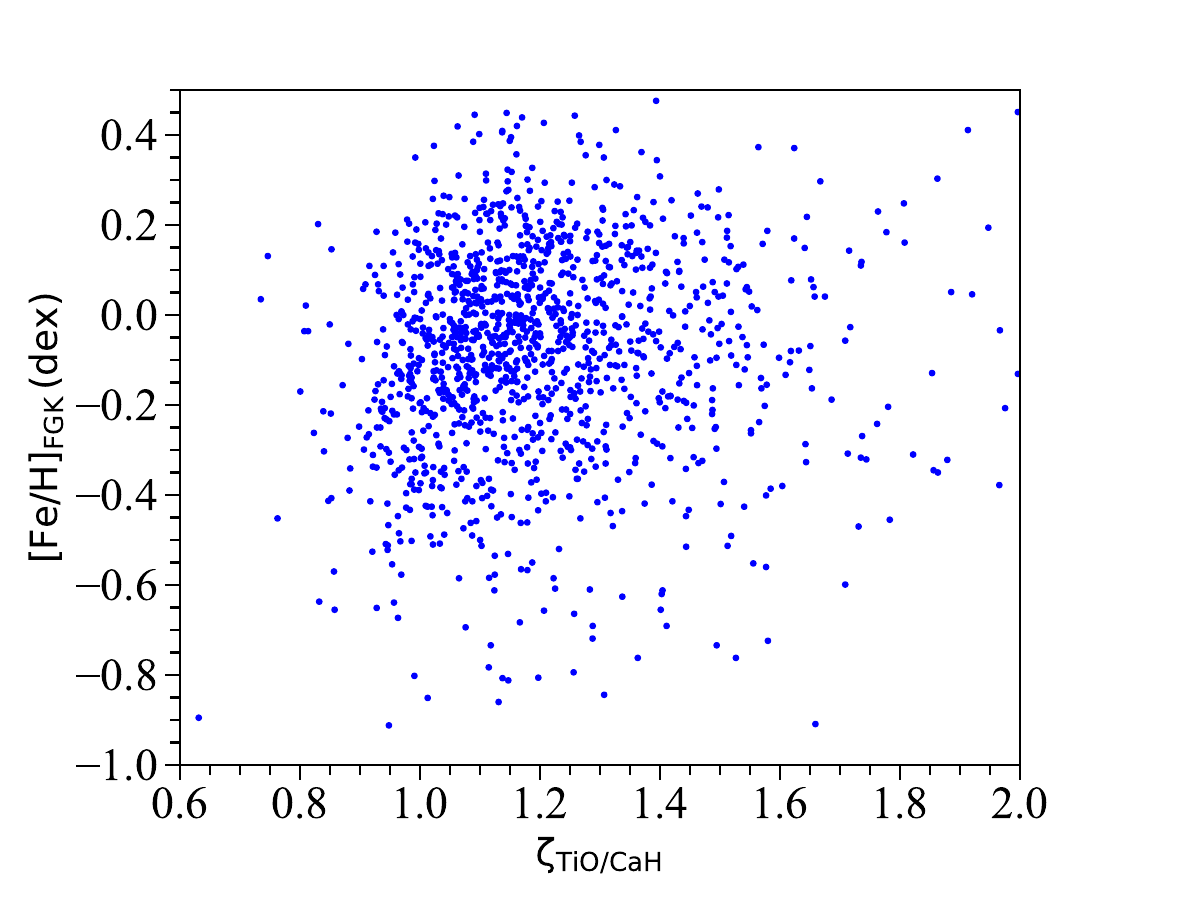}
\caption{The distribution of [Fe/H] and $\zeta\rm_{TiO/CaH}$ of 1308 M dwarfs, the [Fe/H] comes from F, G, or K dwarf companions. }\label{fig:TiO_CaH}
\end{figure}

\citet{Gizis-1997} introduced a classification system of subdwarfs, which is based on the measurements of the four spectroscopic indices, CaH1, CaH2, CaH3 and TiO5. These indices were originally defined by \citet{Reid-1995}. \citet{Lepine-2003} pointed out that they are able to discriminate dwarfs, subdwarfs, and extreme subwarfs of M stars \citep {Reid-2005,Lepine-2007,Hejazi-2020,Zhang-2021,Hejazi-2022}. \citet{Lepine-2007} proposed the $\zeta_{\rm TiO/CaH}$ metallicity index to redefine the metallicity subclass based on the calibration of the TiO to CaH ratio for stars at solar metallicity. $\zeta$ is defined as
\begin{equation}\label{eq:zata}
\zeta\rm_{TiO/CaH}=\frac{1-TiO5}{1-[TiO5]\rm_{{M}_\odot}}.
\end{equation}
TiO5$\rm_{{M}_\odot}$ is a cubic polynomial fit of the TiO5 spectral index as a function of the CaH2+CaH3 index. It effectively provides the calibration of TiO band strength relative to CaH (CaH2+CaH3) band. As shown in below
\begin{equation}\label{eq:sun}
\rm \left[TiO5\right]\rm_{{M}_\odot}=a(CaH)^{3}+b(CaH)^{2}+c(CaH)+d.
\end{equation}
In this work, we take the coefficients provided by \citet{Lepine-2013}, where a, b, c, and d are -0.588, 2.211, -1.906, and 0.622, respectively. Stars with $\zeta$ $>$0.825, 0.825$>$ $\zeta$ $>$0.5, 0.5$>$ $\zeta$ $>$0.2 and $\zeta$ $<$0.2 are classified as M dwarfs (dM), subdwarfs (sdM), extreme-subdwarfs (esdM), and ultra-subdwarfs (usdM), respectively. Figure \ref{fig:TiO_CaH} shows the distribution of \feh\ and $\zeta\rm_{TiO/CaH}$ for the 1308 M dwarfs selected in subsection \ref{sect:binary}. Except that the $\zeta\rm_{TiO/CaH}$ index of 10 stars with low spectral signal-to-noise ratio is less than 0.825, the other training samples in this work have $\zeta\rm_{TiO/CaH}$ $>$0.825, which indicates that most of the training samples in this work are M dwarfs. 

\citet{Woolf-2009} demonstrated that there is a linear correlation between $\zeta\rm_{TiO/CaH}$ and \feh\ in M stars. \citet{Lepine-2013} also found a weak correlation between $\zeta$ and \feh\ with 0.9$<$ $\zeta\rm_{TiO/CaH}$ $<$ 1.2. \citet{Mann-2013a} proposed that $\zeta$ is correlated with \feh\ for stars only at super-solar metallicity, but not at low metallicity. In Figure \ref{fig:TiO_CaH}, we show the same trend between $\zeta\rm_{TiO/CaH}$ and \fehfgk, similar to the Figure 16 of \citet{Lepine-2013}. We obtained that the Pearson correlation coefficient and the two-tailed $p$-value associated with the Pearson correlation coefficient between $\zeta\rm_{TiO/CaH}$ and \fehfgk\ are 0.12 and 1.2e-05, respectively. It demonstrates a moderate linear correlation between $\zeta\rm_{TiO/CaH}$ and \feh\ for the $\sim$1300 M dwarfs in the range of -1 $<$[Fe/H]$<$ 0.5 in this work. This is expected as the TiO and CaH features not only depend on the metallicity but also have an important correlation with other parameters such as $\alpha$-element enhancement [$\alpha$/Fe]. 

\subsection{Metallicity analysis}\label{sect:Meta}
\begin{figure*}
\centering
\includegraphics[width=1.1\textwidth, trim=3.5cm 0.0cm 0.0cm 0.cm, clip]{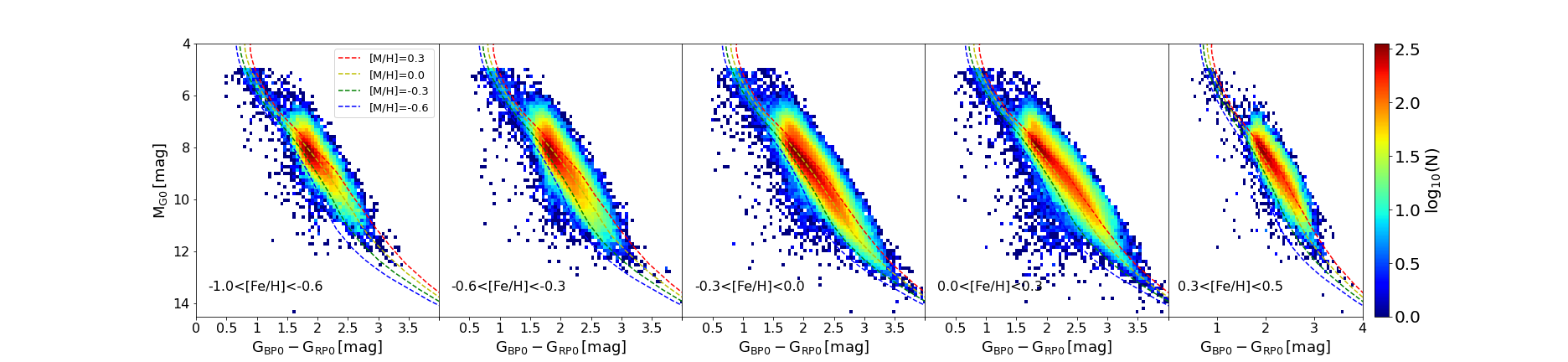}
\includegraphics[width=1.1\textwidth, trim=3.5cm 0.0cm 0.0cm 0.cm, clip]{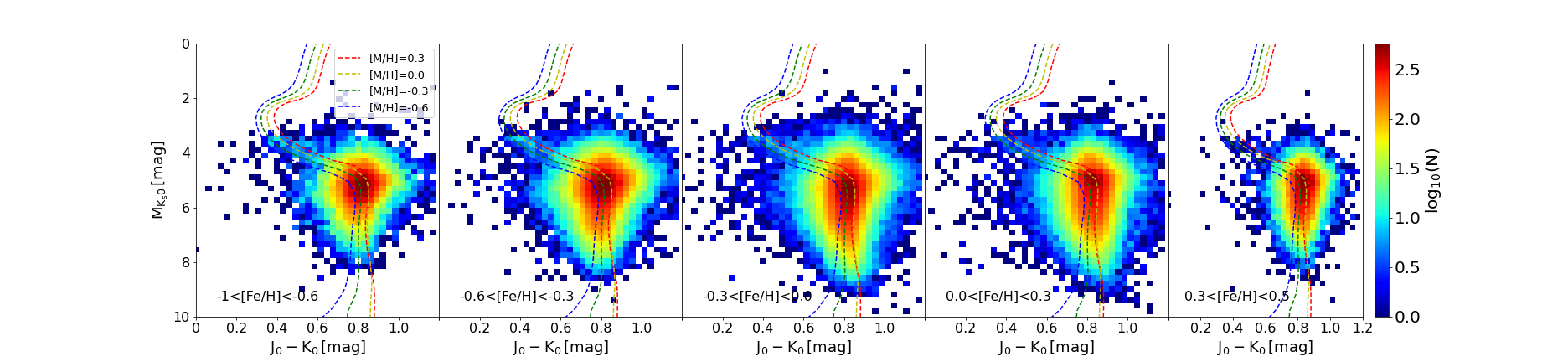}
\caption{The top five panels display the $M\rm_{BP0}-M\rm_{RP0}$ vs. $M\rm_{G0}$ diagrams of M dwarfs in different metallicity bins. The colors are coded by the logarithm of the number of M dwarfs in each $M\rm_{BP0}-M\rm_{RP0}$ and $M\rm_{G0}$ bins. The four dashed lines in each panel are same as those described in Figure \ref{fig:BP_RP_tr}. The bottom five panels are similar to the top panels, but in $J\rm_{0}-K\rm_{0}$ vs. $M\rm_{K\rm_{s}0}$ plane. Each star has been corrected for extinction.}\label{fig:feh_bin}
\end{figure*}

About 630,000 M dwarfs are divided into five metallicity bins. The  ($M\rm_{BP0}-M\rm_{RP0}$ vs. $M\rm_{G0}$) diagram of these five subsets are shown in the five top-panels of Figure \ref{fig:feh_bin}. Each star has been corrected for extinction as described in subsection \ref{sect:Proper}. The colors are encoded by the logarithm of the number of stars in each $M\rm_{BP0}-M\rm_{RP0}$ and $M\rm_{G0}$ bins. It is obvious that the M dwarfs with \feh $>$ -0.3 dex are consistent with the isochrones \citep{Bressan-2012}. However, there is a deviation between predicted metallicity and the corresponding isochrones in the range of -1 $<$\feh$<$ -0.3 dex. The five bottom panels are similar to the five top panels. They display the $J\rm_{0}-K\rm_{0}$ vs. $M\rm_{K\rm_{s}0}$ diagrams of the five subsets. The extinction coefficients of the $J$ and $K$ bands are from \citet{Wang-2019}, where $A_J$= 0.243*$A_V$ and $A_K$= 0.078*$A_V$. The five \feh\ bins are roughly consistent with the corresponding isochrones in near-infrared bands, especially for M dwarfs with \feh$>$ -0.3 dex. However, it cannot be ignored that there are still differences between the PARSEC model and the predicted metallicity in the range of \feh$<-0.6$ dex. Considering that the stars with low metallicity are mostly old stars, the stellar activity does not have a great impact on photometric measurement. We point out that the PARSEC model of M dwarfs with low metallicity is inconsistent with the results of our work, which may be due to insufficient understanding of continuous opacities of M dwarfs. Alternatively, some other factors like the method, low-resolution spectra and wavelength ranges used in our work also may contribute to the difference. 

\section{Conclusion}\label{sect:conclu}

In this work, we identified 1308 FGK+M wide binary systems. We calibrated the \feh\ of the M dwarfs from their F, G, or K companions. The \feh\ of the M dwarfs is in the range from -1 to 0.5 dex, and effective temperature spanning in 3100 $<$ \teff\ $<$ 4400 K. By training a data-driven model SLAM based on the 1308 M dwarf secondaries, we derived the stellar parameters (\feh\ and \teff) for $\sim$ 630,000 LAMOST M dwarf stars with low-resolution optical spectra. The precision of \feh\ and \teff\ are 0.15\,dex and 40\,K at $snri$=100, respectively. 

We used two methods to verify the self consistency of the stellar parameters determined from the SLAM model. The first one is dividing the 1308 FGK+M wide binaries into the training and test set. The bias and scatter of metallicty and temperature of the test data are 0.01$\pm$0.19 dex and 2$\pm$54 K, respectively. In the second method, we chose 606 M+M wide binaries, the \feh\ of the two components in each wide binary are derived from the SLAM model independently. The bias value of the two components of the 606 M+M wide binaries is 0.02 dex with a scatter of 0.15 dex. Both methods indicate that the atmospheric parameters determined by the SLAM model are precise. 

We compared our resulting \feh\ and \teff\ values with the literature. For \feh, there is near zero bias with a scatter of  $\sim$ 0.14-0.18 dex compared to \citet{Birky-2020}, who also calibrated \feh\ using F, G, and K companions. However, \fehslam\ are systematically higher than \fehap\ with an offset of  0.10$\pm$0.15 dex to 0.15$\pm$0.20 dex. This systematic difference may be caused by the different atomic or molecular lines, or uncertainties in the continuous opacity of the stellar atmospheric model used by APOGEE pipeline. Many other factors like different spectral resolution, different wavelength ranges and different methods used in our work from those used in APOGEE may also contribute to the systematic difference. Compared to the temperature calibrated by bolometric temperature, the \teffslam\ is overestimated by 180 K. But there is a good consistency between our temperature and that of APOGEE.

We calculated $\zeta$ for the 1308 M dwarf secondaries. It is originally defined for the classification of M stars. The Pearson correlation coefficient between $\zeta$ and \fehfgk\ is only 0.12, which indicates that there is a moderate correlation between these two parameters. 

Compared to LAMOST, the upcoming SDSS-V \citep{Kollmeier-2017,Almeida-2023} has the capability to detect fainter stars, making it a promising tool to supplement the lack of metal-poor wide binaries in this study. Our method and catalog will serve as valuable references for deriving fundamental parameters and metallicity within the framework of SDSS-V.


\section{Acknowledgements}
This work is supported by the National Key R\&D Program of China No. 2019YFA0405500 and the China Manned Space Project with no. CMS-CSST-2021-A08. HJT. thanks National Natural Science Foundation of China (NSFC) under grant number 12373033. ZXN acknowledges support from the NSFC through grant No.~12303039.
Guoshoujing Telescope (the Large Sky Area Multi-Object Fiber Spectroscopic Telescope LAMOST) is a National Major Scientific Project built by the Chinese Academy of Sciences. Funding for the project has been provided by the National Development and Reform Commission. LAMOST is operated and managed by the National Astronomical Observatories, Chinese Academy of Sciences.
This work used the data from the European Space Agency (ESA) mission Gaia (https://www.cosmos.esa.int/gaia), processed by the Gaia Data Processing and Analysis Consortium (DPAC; https://www.cosmos.esa.int/web/gaia/dpac/consortium). Funding for the DPAC has been provided by national institutions, in particular the institutions participating in the Gaia Multilateral Agreement.

Facilities: LAMOST, Gaia.

Software: astropy \citep{Astropy-2018}, scipy \citep{Virtanen-2020},
scikit-learn \citep{Pedregosa-2012},SLAM \citep{Zhang-2020}, TOPCAT \citep{Taylor-2005}.

\section*{Appendix}
\begin{figure*}
\centering
\includegraphics[width=1.0\textwidth, trim=0.0cm 0.0cm 0.0cm 0.cm, clip]{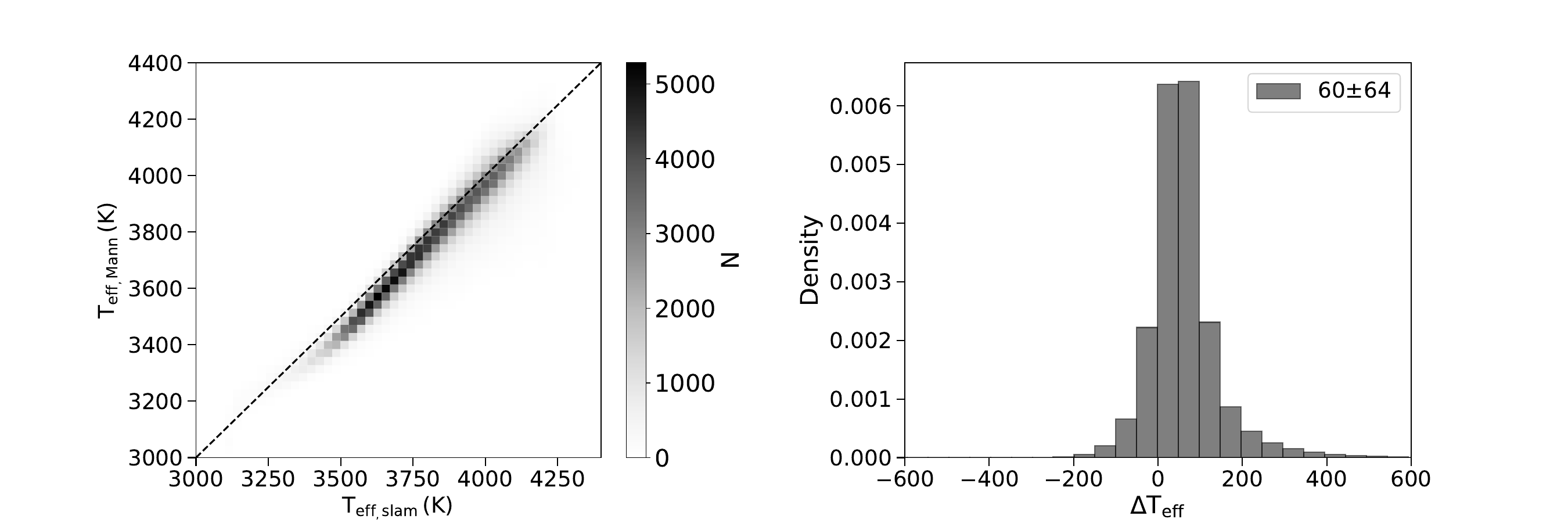}
\caption{The comparison of \teffslam\ and \teffmann\ of stars with \fehslam\ $>$ -0.6 dex, where \teffslam\ and \teffmann\ represent the temeprature of M dwarfs derived from SLAM model and from the empirical relationship provide by \citet{Mann-2016}, respectively. }
\label{fig:teff_slam_mann}
\end{figure*}

In order to ensure that the training and prediction sets in this work are indeed M dwarfs, we derived the photometric temperature (\teffmann) and surface gravity (\loggmann) for stars in our work. The \teffmann of samples calculated from the metallicity-independent empirical relationship provided by \citet{Mann-2015,Mann-2016} between \teff\ and magnitudes from Two Micron All Sky Survey \citep[2MASS][]{Skrutskie-2006} and Gaia, i.e., J, H, BP and RP bands. This relationship is valid for stars with $0.1\, R\odot\,<R_{*}<0.7\, R \odot$ and -0.6 dex $<$ \feh $<$ +0.5 dex, where $R_{*}$ is the stellar radius. The detail coefficients of the relationship we used as described in Table2 of \citet{Mann-2016} corresponding to the one that shows a scatter of 49 K. The left panel of Figure \ref{fig:teff_slam_mann} shows the comparison of \teffmann\ and \teffslam\ of stars with \fehslam $>$ -0.6 dex. It indicates that there has a systematic bias of 60 K with a scatter of 64 K as shown in the corresponding distribution of  $\Delta$ \teff (=\  \teffslam- \teffmann\,) in the right panel. 

\begin{figure}
\centering
\includegraphics[width=0.45\textwidth, trim=0.0cm 0.0cm 0.0cm 0.cm, clip]{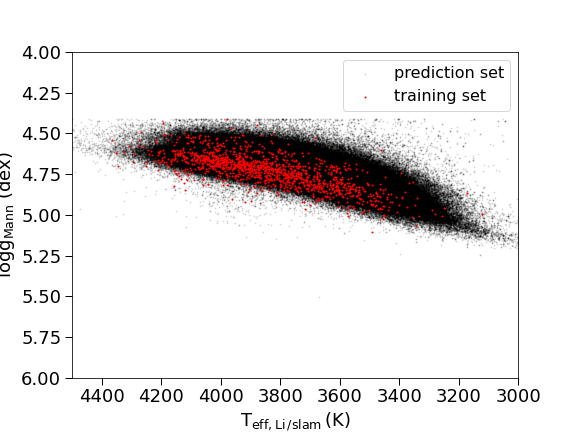}
\caption{The red and black dots exhibit the distribution of temperature and surface gravity of training set (\teffli\, and \loggmann) and prediction sample (\teffslam and \loggmann), respectively.  }
\label{fig:teff_logg}
\end{figure}

The photometric surface gravity of M dwarfs in this work was computed by using the following relation
\begin{equation}\label{eq:logg}
log\,g = 4.438+log_{10}(M_{*}/M_{\odot})-2*log_{10}(R_{*}/R_{\odot})
\end{equation}
where $R_{\odot}$ and $M_{\odot}$ are the stellar radius and mass, respectively. $M_{*}$ is the stellar mass. We inferred the $R_{*}/R_{\odot}$ of M dwarfs by adopting the relationship between $R_{*}/R_{\odot}$ and absolute magnitude in K band as descried in Table 1 of \citet{Mann-2016}, the one that shows a $\sigma$\, of 2.89\%. The stellar mass determined by the relationship between $M_{*}/M_{\odot}$ and the absolute magnitude in K band provided in \citet{Mann-2019}. We used the coefficients of the relationship that shows a Bayesian Information Criterion (BIC) of 86 in Table 6 in \citet{Mann-2019} to derive the mass of M dwarfs in this work. Then \loggmann can be determined according to equation (\ref{eq:logg}). Figure \ref{fig:teff_logg} displays the distribution of \teffslam (\teffli) and \loggmann\ of training sample (red) and the prediction M dwarfs (black). It demonstrates that the \teff\ and \logg\ of all the samples are associated with late-type K and M dwarf stars.

\section*{Data Availability}
The M dwarf parameters in our study are fully public and can be accessed via the link \url{https://nadc.china-vo.org/res/r101265/}.

\bibliographystyle{mnras}
\bibliography{mainNotes.bib}




\begin{table*} 
 \centering
    \caption{Catalog description of $\sim$ 750,000 M dwarfs} 
    \label{col:all}
\begin{tabular}{p{4.0cm}|p{1.5cm}|p{8.0cm}}
\hline\hline
Column & units & Description \\ \hline
{obsid}                          &         & LAMOST DR9 observe id   \\
{source\_id}                          &         & Gaia eDR3 source id   \\
{ra\_obs}                                  & deg     & right ascension from LAMOST  \\
{dec\_obs}                                 & deg     & declination from LAMOST \\
{snru}                    & mag            & signal to noise ratio of u magnitude \\
{snrg}                    & mag            &  signal to noise ratio of g magnitude \\
{snrr}                    & mag            &  signal to noise ratio of r magnitude \\
{snri}                    & mag            &  signal to noise ratio of i magnitude \\
{snrz}                    & mag            &  signal to noise ratio of z magnitude \\
{parallax}                            & mas     & parallax  \\
{parallax\_over\_error}               &         & parallax divided by its error \\
{\fehslam}               & dex     & [Fe/H] from SLAM \\
{\feh$\_{err}$}               & dex     & [Fe/H] uncertainty\\
{\teffslam}               & K     & \teff\ from SLAM \\
{\teff$\_{err}$}               & K     & \teff\ uncertainty\\

\hline
\hline
\end{tabular}
\end{table*}

\end{document}